\documentclass[
  prb,
  reprint
  ]{revtex4-2}

\usepackage{graphicx}
\usepackage{amsmath}
\usepackage{braket}
\usepackage{bm}
\usepackage{xcolor}
\usepackage{hyperref}
\usepackage{subfig}
\usepackage{threeparttable}
\usepackage{booktabs}
\usepackage{chemformula}

\newcommand{\V}[1]{\bm{#1}}

\begin{document}

\title{Scaling up the transcorrelated density matrix renormalization group}

\author{Benjamin Corbett}
  \email[]{bcorbett@unm.edu}
\author{Akimasa Miyake}
  \email[]{amiyake@unm.edu}
\affiliation{Center for Quantum Information and Control, Department of Physics and Astronomy,
University of New Mexico, Albuquerque, New Mexico 87106, USA}

\date{\today}

\begin{abstract}
Explicitly correlated methods, such as the transcorrelated method which shifts a Jastrow or Gutzwiller correlator from the wave function to the Hamiltonian, are designed for high-accuracy calculations of electronic structures, but their application to larger systems has been hampered by the computational cost. We develop improved techniques for the transcorrelated density matrix renormalization group (DMRG), in which the ground state of the transcorrelated Hamiltonian is represented as a matrix product state (MPS), and demonstrate large-scale calculations of the ground-state energy of the two-dimensional Fermi-Hubbard model. Our developments stem from three technical inventions: (i) constructing matrix product operators (MPO) of transcorrelated Hamiltonians with low bond dimension and high sparsity, (ii) exploiting the entanglement structure of the ground states to increase the accuracy of the MPS representation, and (iii) optimizing the non-linear parameter of the Gutzwiller correlator to mitigate the non-variational nature of the transcorrelated method. We examine systems of size up to $12 \times 12$ lattice sites, four times larger than previous transcorrelated DMRG studies, and demonstrate that transcorrelated DMRG yields significant improvements over standard non-transcorrelated DMRG for equivalent computational effort. Transcorrelated DMRG reduces the error of the ground state energy by $2.4\times$~-~$14 \times$, with the smallest improvement seen for a small system at half-filling and the largest improvement in a dilute closed-shell system.
\end{abstract}

\maketitle

\section{Introduction}

Calculating ground state properties of strongly correlated systems of electrons is a central challenge in condensed matter and quantum chemistry. Because the associated Hilbert space grows exponentially with the number of electrons, exact solutions are only possible for the smallest systems. As such, many techniques have been developed over the decades to approximate the ground state such as Hartree-Fock, density functional theory, coupled cluster and quantum Monte Carlo \cite{Hartree-1928, Kohn-Sham-1965, Taylor-1994, Sugiyama-Koonin-1986}. Explicitly correlated methods are another such technique, in which the wave function is augmented by the action of a many-body correlator. First used almost a century ago by Hylleraas to study Helium \cite{Hylleraas-1928, Hylleraas-1930}, explicitly correlated methods have recently experienced a resurgence due to their high accuracy at a surmountable cost \cite{Klopper-Manby-Ten-No-Valeev-2006, Shiozaki-Hirata-2010, Kong-Bischoff-Valeev-2012, Mitroy-Bubin-Horiuchi-2013, Gruneis-Hirata-Ohnishi-2017, Ma-Werner-2018, Schleich-Kottmann-Aspuru-Guzik-2022}. In the transcorrelated method, first introduced by Boys and Handy \cite{Boys-Handy-1969a, Boys-Handy-1969b, Boys-Handy-1969c, Handy-1973}, the correlator is moved from the wave function onto the Hamiltonian by means of a similarity transformation, producing an effective transcorrelated Hamiltonian. In most cases, the correlator is not a unitary operator, and hence the transcorrelated Hamiltonian is not Hermitian. Because of this, the variational principle does not apply, and it is possible to obtain expectation values below the ground state energy.

In the initial applications of the transcorrelated method, the ground state of the transcorrelated Hamiltonian was approximated by a single Slater determinant \cite{Boys-Handy-1969a, Boys-Handy-1969b, Boys-Handy-1969c, Handy-1973, Ten-no-2000, Tsuneyuki-2008, Neuscamman-Changlani-Kinder-2011, Wahlen-Strothman-Hoyos-2015}. More recently, transcorrelated ground states have been approximated by full configuration interaction quantum Monte Carlo (FCIQMC) ensembles \cite{Luo-Alavi-2018, Dobrautz-Luo-Alavi-2019, Cohen-Luo-2019, Jeszenszki-Ebling-Luo-2020, Haupt-Hosseini-Mohammadreza-2023}, matrix product states (MPS) \cite{Baiardi-Reiher-2020, Baiardi-Lesiuk-Reiher-2022, Liao-Zhai-Christlmaier-2023} and states prepared on quantum computers \cite{McArdle-Tew-2020, Motta-Gujarati-Rice-2020, Kumar-Asthana-2022, Sokolov-Dobrautz-Luo-2023, Magnusson-Fitzpatrick-Knecht-2024, Dobrautz-Sokolov-Liao-2024}. MPSs in particular are an apt choice. Enabled by the density matrix renormalization group (DMRG), a classical algorithm that optimizes the MPS approximation to the ground state of a Hamiltonian \cite{White-1992, White-1993}, MPSs are now the default tool for studying one dimensional systems \cite{Schollwock-2005, White-Affleck-2008, Barthel-Schollwock-White-2009, White-2009}. In principle any state can be represented exactly as an MPS although with a bond dimension (a measure of complexity) that grows exponentially with the system size. But, as shown in Ref.~\cite{Schuch-Wolf-Verstraete-2008}, certain low entanglement states have an accurate approximation with a polynomial bond dimension. These properties make MPSs an enticing partner of the transcorrelated method, which should reduce the correlations in the ground state, making it more suitable MPS approximation.

In the case of the electronic structure Hamiltonian, the correlator is often chosen to satisfy Kato's cusp condition, which describes the behavior of the eigenstates near the coalescence point of two electrons \cite{Kato-1957}. In fact, the success of Hylleraas's method on Helium can be attributed to satisfying this cusp condition three decades before its discovery. For the Fermi-Hubbard Hamiltonian \cite{Gutzwiller-1963, Hubbard-1963}, a common choice is the Jastrow correlator (or the Gutzwiller correlator in the second quantized picture) \cite{Jastrow-1955, Gutzwiller-1963, Brinkman-Rice-1970, Ogawa-Kanda-Matsubara-1975, Bunemann-Weber-Gebhard-1998} which has only one parameter, but unfortunately there is no corresponding condition to fix its value.

Following the seminal work of Baiardi and Reiher in Ref.~\cite{Baiardi-Reiher-2020}, we apply the transcorrelated method to the momentum space representation of the two-dimensional Fermi-Hubbard Hamiltonian using the Gutzwiller correlator and approximate the ground state as an MPS which we then optimize using DMRG. Derived from the electronic structure Hamiltonian of a crystal through significant approximations \cite{Essler-Frahm-2005}, the Fermi-Hubbard Hamiltonian provides a valuable theoretical tool for analyzing the properties of correlated electronic phenomena such as the Mott metal-insulator transition \cite{Villagran-Mitsakos-Lee-2020}, superconductivity \cite{Qin-Chung-Shi-2020} and charge-spin density waves \cite{Peters-Kawakami-2014}. In one dimension, the Fermi-Hubbard Hamiltonian is solvable via the Bethe ansatz \cite{Lieb-Wu-1968}; however, in two or more dimensions, it remains unsolved, and accurately determining properties of the ground state poses a major computational challenge. The ground state of the one and quasi-two-dimensional Fermi-Hubbard Hamiltonians can be accurately approximated as small-bond-dimension MPSs \cite{White-Scalapino-2000, White-Scalapino-2003, Ehlers-White-Noack-2017}. The ground state of the two-dimensional system however, has long range correlations when mapped to the one-dimensional MPS, prohibiting accurate approximations at small bond dimensions. One benefit of studying the Fermi-Hubbard Hamiltonian is that we have reference energies calculated using auxiliary field quantum Monte Carlo (AFQMC) in Refs.~\cite{Shi-Zhang-2013,Qin-Shi-Zhang-2016}. At half-filling, due to the particle-hole symmetry, the AFQMC calculations are sign-problem free and the authors look at systems with up to 256 sites. Away from half-filling, however, the sign problem introduces an exponential systematic error into AFQMC results \cite{Loh-Gubernatis-1990, Shi-Zhang-2016}. Nevertheless, through the use of symmetry, the authors mitigate the sign problem and claim essentially exact results up to 144 sites.

Due to the three electron terms present in the transcorrelated Hamiltonian, constructing its matrix product operator (MPO) representation limits the systems we are able to explore. Using a highly optimized and optimal construction algorithm, we examine systems with up to 144 sites. While not as large as the systems examined with AFQMC, it is still a significant improvement over the previous transcorrelated approaches which were limited to 50 sites with FCIQMC \cite{Dobrautz-Luo-Alavi-2019} and 36 sites with DMRG \cite{Baiardi-Reiher-2020}. By recognizing and exploiting a pattern in the entanglement structure of the ground state, we are able to construct new mappings from momentum modes to MPS tensors which both improve the accuracy of the DMRG calculations and facilitate comparisons between different systems. Additionally, we develop a framework to optimize the correlator along with the MPS in a self-consistent manner, which mitigates the non-variational nature of the transcorrelated method. Unlike both prior approaches, which used a fixed correlator, we never obtain energies below the reference energies. While we are unable to match the accuracy of the AFQMC calculations, we aim to advance the field of transcorrelated DMRG calculations and illuminate the method's potential. The rest of the paper is organized as follows. First we introduce the Fermi Hubbard Hamiltonian, and the transcorrelated method in Sec.~\ref{sec:Transcorrelated Fermi-Hubbard} followed by a brief review of matrix product states and associated algorithms in Sec.~\ref{sec:Matrix Product States}. We present the technical inventions outlined above in Sec.~\ref{sec:Technical inventions} and our results in Sec.~\ref{sec:Results}. Finally, we summarize our work and present the outlook in Sec.~\ref{sec:Conclusion}.

\section{Transcorrelated Fermi-Hubbard} \label{sec:Transcorrelated Fermi-Hubbard}

\subsection{The Fermi-Hubbard Hamiltonian}

The Fermi-Hubbard Hamiltonian on $N$ sites is given by
\begin{equation} \label{real space FH}
  \hat{H}_\text{rs} = -\sum_{j, \ell = 1}^N \sum_{\sigma \in \set{\uparrow, \downarrow}} t_{j \ell} \hat{a}^\dagger_{j, \sigma} \hat{a}_{\ell, \sigma} + U \sum_{j = 1}^N \hat{n}_{j, \uparrow} \hat{n}_{j, \downarrow} ,
\end{equation}
where $\hat{a}^\dagger_{j, \sigma}(\hat{a}_{j, \sigma})$ is the fermionic creation(annihilation) operator at site $j$ with spin $\sigma$ and $\hat{n}_{j, \sigma} = \hat{a}^\dagger_{j, \sigma} \hat{a}_{j, \sigma}$ is the number operator. The first sum is called the ``hopping'' term, which we denote as $\hat{T}$, and describes electrons hopping between sites with amplitude given by the Hermitian matrix $t_{j \ell}$. The second sum is called the ``interaction'' term, which we denote as $\hat{D}$, and describes the onsite repulsion between electrons; in this repulsive regime $U \geq 0$. Note that $\hat{D}$ counts the number of doubly occupied sites. With this convention the Hamiltonian can be compactly written as $\hat{H} = -\hat{T} + U \hat{D}$. An important special case is when one can map the sites to a finite $d$-dimensional lattice such that $t_{j \ell} = f(\V{r}_j - \V{r}_\ell)$ where $\V{r}_j$ are the coordinates of the $j$-th site and $f$ is a periodic function. In this case, the Hamiltonian is invariant under translation by a lattice vector. By performing a single electron basis rotation we can diagonalize the hopping term. This is done via the following unitary transformation
\begin{align} \label{eq:rs in terms of ks}
  \hat{a}_{j, \sigma} &= \frac{1}{\sqrt{N}} \sum_{\V{k} \in G} e^{-i \V{k} \cdot \V{r}_j} \hat{c}_{\V{k}, \sigma} ,
\end{align}
where $G$ is the reciprocal lattice. After the substitution the Hamiltonian is
\begin{equation} \label{eq:momentum space FH}
  \hat{H}_\text{ks} = - \sum_{\substack{\V{k} \in G \\ \sigma}} \epsilon(\V{k}) \hat{n}_{\V{k}, \sigma} + \frac{U}{N} \sum_{\V{p}, \V{q}, \V{k} \in G} \hat{c}^\dagger_{\V{p} - \V{k}, \uparrow} \hat{c}^\dagger_{\V{q} + \V{k}, \downarrow} \hat{c}_{\V{q}, \downarrow} \hat{c}_{\V{p}, \uparrow}
\end{equation}
where $\epsilon$ depends on the form of $f$. We consider square lattices with periodic boundary conditions and nearest neighbor hopping such that
\begin{equation} \label{eq:epsilon}
  \epsilon(\V{k}) = 2 t \left( \cos k_x + \cos k_y  \right) \ ,
\end{equation}
where $\V{k} = (k_x, k_y)$ and $k_x, k_y \in \frac{2 \pi}{L} [1, L]$ for a real space grid of size $L \times L$. In this context Eq.~\eqref{real space FH} is called the real space Hamiltonian and Eq.~\eqref{eq:momentum space FH} is called the momentum space Hamiltonian.

The Fermi-Hubbard Hamiltonian has many symmetries, the most relevant of which are the conservation of the total number of electrons $\hat{N}_e = \sum_{j \sigma} \hat{n}_{j, \sigma}$, conservation of the total z component of the spin $\hat{S}_z = \sum_{j} (\hat{n}_{j, \uparrow} - \hat{n}_{j, \downarrow}) / 2$, and the translation invariance in real space which becomes conservation of total momentum modulo $2 \pi$ in momentum space
\begin{equation} \label{total momentum}
  \hat{\V{K}} = \sum_{\substack{\V{k} \in G \\ \sigma}} \V{k} \hat{n}_{\V{k}, \sigma} \mod 2 \pi .
\end{equation}
All of our calculations are carried out in the $S_z = \V{K} = 0$ sector, and an important quantity will be the filling factor $\eta = N_e / (2 N)$. Half-filling is when $\eta = 0.5$, and dilute systems are those for which $\eta < 0.5$.

\subsection{The Transcorrelated Method}

In the transcorrelated method, instead of representing the ground state of the Hamiltonian $\ket{\psi}$, we instead choose to represent $\ket{\phi}$ such that 
\begin{equation}
  \ket{\psi} = e^{\hat{\tau}} \ket{\phi} ,
\end{equation}
for some parameterized operator $\hat{\tau}$ known as the correlator. In order to solve for $\ket{\phi}$ we multiply the standard eigenvalue equation by $e^{-\hat{\tau}}$ from the left to get
\begin{equation}
  e^{-\hat{\tau}} \hat{H} e^{\hat{\tau}} \ket{\phi} = \widetilde{H} \ket{\phi} = E \ket{\phi} ,
\end{equation}
where $\widetilde{H}$ is the transcorrelated Hamiltonian. The idea is to choose the correlator such that the ground state of the transcorrelated Hamiltonian, $\ket{\phi}$, is easier to represent or approximate than the ground state of the initial Hamiltonian, $\ket{\psi}$. Except in the specific case when $\hat{\tau}$ is anti-Hermitian, the transcorrelated Hamiltonian will no longer be Hermitian and will have different left and right eigenvectors. The main consequence of this is if $\ket{\phi}$ is not an exact eigenstate the ground state energy is no longer a lower bound on the transcorrelated energy, $E = \braket{\phi | \widetilde{H} | \phi}$.

\subsection{Transcorrelated Fermi-Hubbard}

Here we use the Gutzwiller correlator, $\hat{\tau} = J \hat{D}$ for real valued $J \leq 0$. Unlike the transcorrelated electronic structure Hamiltonian, which is usually constructed in first quantization and then quantized, because the Gutzwiller correlator is already in second quantized form, the transcorrelated Hamiltonian is isospectral to the Hamiltonian and does not suffer from errors introduced from a finite basis set. The transcorrelated Hamiltonian can be derived via the Baker-Campbell-Hausdorff formula as done in Ref.~\cite{Dobrautz-Luo-Alavi-2019}, below we derive it via a different approach.

Consider a resolution of the identity $\bar{n}_{j, \sigma} \equiv \hat{1} - \hat{n}_{j, \sigma}$. As $\hat{n}_{j, \sigma}$ is the projector onto the subspace in which site $j$ is occupied by a spin-$\sigma$ electron, $\bar{n}_{j, \sigma}$ is the projector onto the subspace in which site $j$ is not occupied by a spin-$\sigma$ electron. We can use this identity to rewrite the hopping term as
\begin{align}
  \hat{T} &= \sum_{j, \ell, \sigma} t_{j \ell} \bar{n}_{j, \bar{\sigma}} \hat{a}^\dagger_{j, \sigma} \hat{a}_{\ell, \sigma} \bar{n}_{\ell, \bar{\sigma}} + \sum_{j, \ell, \sigma} t_{j \ell} \hat{n}_{j, \bar{\sigma}} \hat{a}^\dagger_{j, \sigma} \hat{a}_{\ell, \sigma} \hat{n}_{\ell, \bar{\sigma}} \nonumber \\
  &+ \sum_{j, \ell, \sigma} t_{j \ell} \bar{n}_{j, \bar{\sigma}} \hat{a}^\dagger_{j, \sigma} \hat{a}_{\ell, \sigma} \hat{n}_{\ell, \bar{\sigma}} + \sum_{j, \ell, \sigma} t_{j \ell} \hat{n}_{j, \bar{\sigma}} \hat{a}^\dagger_{j, \sigma} \hat{a}_{\ell, \sigma} \bar{n}_{\ell, \bar{\sigma}} \nonumber \\
  &\equiv \hat{T}_{ss} + \hat{T}_{dd} + \hat{T}_{sd} + \hat{T}_{ds} ,
\end{align}
where $\bar{\sigma}$ is the opposite spin of $\sigma$. This is a particularly useful representation since $\hat{T}_{ss}$ describes an electron leaving a singly occupied site to create a new singly occupied site, $\hat{T}_{sd}$ describes an electron leaving a doubly occupied site to create a new singly occupied site and so on. As such, both $\hat{T}_{ss}$ and $\hat{T}_{dd}$ leave the total number of doubly occupied sites unchanged and hence commute with $\hat{D}$. Now we look at the effect of transcorrelation on $\hat{T}_{sd}$ in the eigenbasis of $\hat{D}$, $\hat{D} \ket{\lambda, j} = \lambda \ket{\lambda, j}$ where $j$ is a label to index the degeneracies in the spectrum, 
\begin{equation}
  \braket{\lambda, j | e^{-J \hat{D}} \hat{T}_{sd} e^{J \hat{D}} | \lambda', j'} = e^{J (\lambda' - \lambda)} \braket{\lambda, j | \hat{T}_{sd} | \lambda', j'} .
\end{equation}
Because $\hat{T}_{sd}$ only includes terms that describe a single electron moving from a doubly occupied site to create a singly occupied site
\begin{equation}
  \braket{\lambda, j | \hat{T}_{sd} | \lambda', j'} = \delta_{\lambda, \lambda' - 1} \braket{\lambda, j | \hat{T}_{sd} | \lambda', j'} ,
\end{equation}
and using this we get
\begin{equation}
  \braket{\lambda, j | e^{-J \hat{D}} \hat{T}_{sd} e^{J \hat{D}} | \lambda', j'} = e^{J} \braket{\lambda, j | \hat{T}_{sd} | \lambda', j'} .
\end{equation}
Since $\{ \ket{\lambda, j} \}_{\lambda, j}$ form a complete basis, and the above equality holds for any pair of basis states
\begin{equation}
  e^{-J \hat{D}} \hat{T}_{sd} e^{J \hat{D}} = e^{J} \hat{T}_{sd} .
\end{equation}
Equivalent logic can be applied to $\hat{T}_{ds}$ to arrive at 
\begin{equation} \label{eq:transcorrelated hopping term}
  e^{-J \hat{D}} \hat{T} e^{J \hat{D}} = \hat{T}_{ss} + \hat{T}_{dd} + e^{J} \hat{T}_{sd} + e^{-J} \hat{T}_{ds} .
\end{equation}

Rewriting the transcorrelated Hamiltonian in a more compact form we arrive at the standard description
\begin{align}
  \widetilde{H}_\text{rs} &= \hat{H}_\text{rs} - \left( e^{J} - 1 \right) \sum_{j, \ell, \sigma} t_{j \ell} \hat{a}^\dagger_{j, \sigma} \hat{a}_{\ell, \sigma} \hat{n}_{\ell, \bar{\sigma}} \nonumber \\
  & -\left( e^{-J} - 1 \right) \sum_{j, \ell, \sigma} t_{j \ell} \hat{n}_{j, \bar{\sigma}} \hat{a}^\dagger_{j, \sigma} \hat{a}_{\ell, \sigma} \nonumber \\
  &+ 2 (\cosh(J) - 1) \sum_{j, \ell, \sigma} t_{j \ell} \hat{n}_{j, \bar{\sigma}} \hat{a}^\dagger_{j, \sigma} \hat{a}_{\ell, \sigma} \hat{n}_{\ell, \bar{\sigma}} .
\end{align}
Transcorrelation roughly triples the number of terms in the Hamiltonian and maintains the geometric locality. However, in real space $e^{J \hat{D}}$ can be represented as an MPO of bond dimension one, therefore $\ket{\psi} = e^{J \hat{D}} \ket{\phi}$ is no harder to represent as an MPS than $\ket{\phi}$. This explains why Ref.~\cite{Baiardi-Reiher-2020} demonstrated essentially no change in the energy by applying the transcorrelated method to the real space Hamiltonian.

By performing the basis rotation, Eq.~\eqref{eq:rs in terms of ks}, we obtain the momentum space transcorrelated Hamiltonian
\begin{align} \label{eq:momentum space transcorrelated Hamiltonian}
  \widetilde{H}_\text{ks} &= \hat{H}_\text{ks} - \sum_{\V{p}, \V{q}, \V{k}, \sigma} \omega(\V{p}, \V{k}) \hat{c}^\dagger_{\V{p} - \V{k}, \sigma} \hat{c}^\dagger_{\V{q} + \V{k}, \bar{\sigma}} \hat{c}_{\V{q}, \bar{\sigma}} \hat{c}_{\V{p}, \sigma} \nonumber \\
  &+ \sum_{\substack{\V{k}, \V{k}' \\ \V{s}, \V{q} \\ \V{p}, \sigma}} \gamma(\V{p}, \V{k}, \V{k}') \hat{c}^\dagger_{\V{p} - \V{k}, \sigma} \hat{c}^\dagger_{\V{q} + \V{k}', \bar{\sigma}} \hat{c}^\dagger_{\V{s} + \V{k} - \V{k}', \bar{\sigma}} \hat{c}_{\V{s}, \bar{\sigma}} \hat{c}_{\V{q}, \bar{\sigma}} \hat{c}_{\mathbf{p}, \sigma} ,
\end{align}
where
\begin{align}
  \omega(\V{p}, \V{k}) &= \frac{1}{N} \left[ (e^J - 1) \epsilon(\V{p} - \V{k}) + (e^{-J} - 1) \epsilon(\V{p}) \right] \\
  \gamma(\V{p}, \V{k}, \V{k}') &= 2 \frac{\cosh(J) - 1}{N^2} \epsilon(\V{p} - \V{k} + \V{k}') .
\end{align}
By grouping like terms together the number of three electron terms can be reduced fourfold, but transcorrelation still increases the number of terms by $O(N^2)$. However, it was demonstrated in Refs.~\cite{Dobrautz-Luo-Alavi-2019} and \cite{Baiardi-Reiher-2020} that transcorrelation simplifies both the configuration-interaction and MPS descriptions of the ground state.

\section{Matrix Product States} \label{sec:Matrix Product States}

What follows is a brief introduction to the matrix product state methods used extensively in this work. There is a wealth of literature on the topic, but for a comprehensive review we point the reader to Refs.~\cite{Schollwock-2011, Paeckel-Kohler-Swoboda-2019}. 

\subsection{Matrix Product States}

For a product state basis $\mathcal{S}$ on the Hilbert space $\mathcal{H} = \bigotimes_{j = 1}^{N} \mathcal{H}_j$ and an arbitrary state
\begin{equation}
  \ket{\psi} = \sum_{\V{s} \in \mathcal{S}} c_{\V{s}} \ket{\V{s}} \ ,
\end{equation}
an MPS description of $\ket{\psi}$ is
\begin{equation} \label{eq:mps definition}
  \ket{\psi} = \sum_{\V{s} \in \mathcal{S}} \sum_{\chi_1 = 1}^{m_1} \dots \sum_{\chi_{N - 1} = 1}^{m_{N - 1}} M_{1, \chi_1}^{s_1} M_{\chi_1, \chi_2}^{s_2} \dots M_{\chi_{N - 1}, 1}^{s_N} \ket{\V{s}} ,
\end{equation}
where $\ket{\V{s}} = \ket{s_1} \otimes \dots \otimes \ket{s_N}$. For a general state, the memory required to store $c_{\V{s}}$ grows exponentially with the system size $N$. By contrast the memory required to store the MPS description Eq.~\eqref{eq:mps definition} appears to be linear in the system size. The catch of course is that the bond dimension $m = \max_j m_j$, usually grows exponentially with the system size. An equivalent, but perhaps more intuitive and useful, MPS description of $\ket{\psi}$ is
\begin{equation} \label{eq:mixed canonical form}
  \ket{\psi} = \sum_{s_j} \sum_{\ell = 1}^{m_{j - 1}} \sum_{r = 1}^{m_j} M^{s_j}_{\ell, r} \ket{\ell} \ket{s_j} \ket{r} ,
\end{equation}
where $\{ \ket{\ell} \}$, the left basis, forms an orthonormal basis for a subspace of $\bigotimes_{\ell = 1}^{j - 1} \mathcal{H}_\ell$. The right basis, $\{ \ket{r} \}$, is similarly defined. An MPS described by Eq.~\eqref{eq:mixed canonical form} is said to be in ``mixed canonical form about site $j$''. Through a series of gauge transformations any MPS can be brought into mixed canonical form about any site. When an MPS is in mixed canonical form about site $j$, the von Neumann bipartite entanglement entropy between $\bigotimes_{\ell = 1}^{j} \mathcal{H}_\ell$ and $\bigotimes_{\ell = j + 1}^{N} \mathcal{H}_\ell$ can be easily calculated
\begin{equation} \label{eq:entanglement entropy definition}
  S(\rho) = - \operatorname{tr}(\rho \log_2{\rho}) = - \sum_{j} \Sigma_j^2 \log_2(\Sigma_j^2) \ ,
\end{equation}
where $\Sigma_j$ are the singular values of the matrix $M_{(\ell, s_j), r}$ formed by reshaping the onsite tensor $M^{s_j}_{\ell, r}$.

\subsection{Density Matrix Renormalization Group}

DMRG is an algorithm to optimize the MPS approximation to the ground state of a Hamiltonian. When the Hamiltonian is Hermitian, DMRG can be interpreted as a greedy energy optimization algorithm, hence the quality of the approximation is determined both by the MPS bond dimension and the ability of DMRG to avoid local minima. Each step of DMRG consists of three parts. First, the MPS is brought into mixed canonical form about site $j$, Eq.~\eqref{eq:mixed canonical form}. Second, the Hamiltonian is projected into the left-onsite-right basis
\begin{equation} \label{eq:projected Hamiltonian}
  \hat{H}^\text{proj}_{(\ell', s_j', r'), (\ell, s_j, r)} = \bra{\ell'} \bra{s_j'} \bra{r'} \hat{H} \ket{\ell} \ket{s_j} \ket{r} \ .
\end{equation}
Finally, the ground state of the projected Hamiltonian is found, and its coefficients are reshaped to form the new onsite tensor $M^{s_i}_{\ell, r}$. The DMRG procedure as described above is known as single-site DMRG, since it optimizes a single site tensor at a time. Throughout this paper, we use the two-site DMRG algorithm which optimizes two adjacent site tensors at once and is less likely to become trapped in local minima \cite{Takahashi-Hikihara-Nishino-1999, White-2005, Schollwock-2011}.

\subsection{Matrix product operators}

As phrased above, DMRG is agnostic as to the representation of the Hamiltonian so long as it can be projected into the left-onsite-right basis. However, when the Hamiltonian is represented as an MPO, DMRG becomes more efficient. For a product operator basis $\mathcal{P}$ for operators on $\mathcal{H}$ and an arbitrary operator
\begin{equation} \label{eq:operator starting point}
  \hat{O} = \sum_{\hat{P} \in \mathcal{P}} c_{P} \hat{P}
\end{equation}
an MPO description of $\hat{O}$ is
\begin{equation} \label{eq:mpo definition}
  \hat{O} = \sum_{\hat{P} \in \mathcal{P}} \sum_{\chi_1 = 1}^{w_1} \dots \sum_{\chi_{N - 1} = 1}^{w_{N - 1}} W_{1, \chi_1}^{P_1} W_{\chi_1, \chi_2}^{P_2} \dots W_{\chi_{N - 1}, 1}^{P_N} \hat{P} ,
\end{equation}
where $\hat{P} = \hat{P}_1 \otimes \dots \otimes \hat{P}_N$. The MPO bond dimension, $w = \max_j w_j$, plays a critical role in the computational complexity of DMRG. Any operator with $k$ nonzero terms $c_P$ can be represented as an MPO of bond dimension $k$. But this representation is not necessarily optimal; for example any operator of the form
\begin{equation} \label{eq:weight one operator}
  \hat{O} = \sum_{j = 1}^N \hat{O}^{(j)} \ ,
\end{equation}
where $\hat{O}^{(j)}$ is an operator that acts non-trivially only on $\mathcal{H}_j$, can be represented as an MPO of bond dimension two.

There are many algorithms to construct an MPO representation of an operator given in the form of Eq.~\eqref{eq:operator starting point} \cite{Keller-Dolfi-Troyer-2015, Hubig-McCulloch-2017, Ehlers-White-Noack-2017, Chan-Keselman-Nakatani-2016, Ren-Li-Jiang-2020}. Of these algorithms two stand out; the rank decomposition approach \cite{Chan-Keselman-Nakatani-2016}, which is guaranteed to produce an MPO of optimal bond dimension but is costly, and the bipartite graph method \cite{Ren-Li-Jiang-2020}, which is more efficient but does not produce an optimal MPO.

The core component of both algorithms is the factorization of a bipartite operator
\begin{equation} \label{eq:two site Hamiltonian}
  \hat{O} = \sum_{a = 1}^{N_A} \sum_{b = 1}^{N_B} \gamma_{ab} \hat{A}_a \otimes \hat{B}_b \ ,
\end{equation}
where $\{ \hat{A}_a \}_a(\{ \hat{B}_b \}_b)$ is a linearly independent set of operators, into a two-site MPO of bond dimension $w$
\begin{equation} \label{eq:rank decomposition ansatz}
  \hat{O} = \sum_{\chi = 1}^w \left( \sum_{a = 1}^{N_A} \alpha_{\chi a} \hat{A}_a \right) \otimes \left( \sum_{b = 1}^{N_B} \beta_{\chi b} \hat{B}_b \right) .
\end{equation}
In the rank decomposition algorithm $\hat{A}_a \otimes \hat{B}_b$ is a product operator in $\mathcal{P}$, and a rank decomposition (i.e. SVD) of $\gamma$ is used to form a MPO with $w = \operatorname{rank}(\gamma)$, the smallest possible bond dimension. On the other hand, the bipartite graph algorithm lets $\hat{A}_a$ be a general operator and views Eq.~\eqref{eq:two site Hamiltonian} as defining a bipartite graph with the operators $\hat{A}_a$ and $\hat{B}_b$ corresponding to vertices connected by an edge if $\gamma_{a b} \neq 0$. The minimum vertex cover of this graph then yields the MPO description where $w$ is the structural rank \cite{Belabbas-Chen-Zelazo-2023} of $\gamma$. The relaxation of the constraint on $\hat{A}_a$ can have quite the impact. Take for example the operator in Eq.~\eqref{eq:weight one operator} which for simplicity we'll assume acts on $N$ qubits. When using the bipartite graph algorithm, we can write this in the form of Eq.~\eqref{eq:two site Hamiltonian} about site $n$ as
\begin{equation}
  \hat{O} = \left( \sum_{j = 1}^{n} \hat{O}^{(j)}_{(n)} \right) \otimes \hat{1}^{\otimes N - n} + \sum_{j = 1}^{N - n} \hat{1}^{\otimes n} \otimes \hat{O}^{(j)}_{(N - n)} ,
\end{equation}
where $\hat{O}^{(j)}_{(n)}$ is an $n$ qubit operator which acts non-trivially only on qubit $j$. Comparing with Eq.~\eqref{eq:two site Hamiltonian}, we see $N_A = 2$ and $N_B = N - n + 1$. Whereas if we used the rank decomposition algorithm we could not group $\left( \sum_{j = 1}^{n} \hat{O}^{(j)}_{(n)} \right)$ together and would have $N_A = n + 1$. This difference amounts to dealing with a $\gamma$ matrix that is either $2 \times O(N)$ or $O(N) \times O(N)$. In general, with the bipartite algorithm $N_A$ can be made proportional to the incoming bond dimension ($w_{n - 1}$ in Eq.~\eqref{eq:mpo definition}). The downside to the bipartite algorithm is that in certain instances the structural rank is larger than the numerical rank, leading to suboptimal bond dimensions.

\section{Technical inventions} \label{sec:Technical inventions}

Our new contributions consist of three parts. First, we wrote an optimized MPO construction algorithm which enables us to apply the transcorrelated method to larger systems than previously possible. Our second innovation is the exploitation the entanglement structure of the ground state of the momentum space Fermi-Hubbard Hamiltonian. We use this pattern to construct two new efficient mappings from momentum space modes to MPS tensors. Finally, we describe a scheme for optimizing the correlator alongside the state $\ket{\phi}$ that not only results in more accurate energies but prevents us from obtaining energies below the ground state energy.

All results were obtained using `ITensors' \cite{ITensors} and `ITensorMPOConstruction' \cite{ITensorMPOConstruction}. Calculations are done with nearest neighbor hopping of uniform amplitude $t = 1$ and interaction strength $U = 4$ on a square lattice with periodic boundary conditions. This choice of $U / t$ corresponds to the intermediate regime where the ground state in both the real and momentum space representations is highly multi reference. We chose a local Hilbert space of dimension four such that we can map the state of each momentum mode $\{ \ket{\emptyset}, \ket{\downarrow}, \ket{\uparrow}, \ket{\downarrow \uparrow} \}$ to a single MPS tensor. The details of our DMRG procedure are given in Appendix \ref{sec:DMRG procedure}.

For all calculations we use a symmetry adapted MPS \cite{Sukhwinder-Pfeifer-Vidal-2011} to fix the number of electrons, total z-spin and total momentum. We set the total z-spin and momentum to zero. For the dilute systems, the AFQMC reference energies were calculated in the $\V{K} = 0$ sector \cite{Shi-Zhang-2013}, but at half-filling momentum conservation was not enforced \cite{Qin-Shi-Zhang-2016}. At half-filling, with both $U$ and $t$ non-zero, the ground state of any bipartite lattice is non-degenerate \cite{Lieb-1989}. For 1D and square 2D periodic systems of even extent $L$, when combined with reflection symmetries this uniqueness implies the ground state has either momentum $\V{K} = 0$ or $\V{K} = \V{\pi}$, where $\V{\pi} = (\pi, \pi)$ in the two-dimensional case. We found numerically that in 1D the ground state had momentum $\pi$ if $L = 0 \mod 4$ and zero otherwise. But in 2D, the ground state of both the $2 \times 2$ and $4 \times 4$ systems had zero momentum. Combined with the reasonable agreement of our energies with the references, we speculate that at half filling the ground state of square periodic lattices has zero momentum.

\subsection{Optimized MPO Construction} \label{sec:Optimized MPO Construction}

Despite the fact that $\hat{H}_\text{ks}$, Eq.~\eqref{eq:momentum space FH}, and $\widetilde{H}_\text{ks}$, Eq.~\eqref{eq:momentum space transcorrelated Hamiltonian}, have $O(N^2)$ and $O(N^5)$ terms for an $N$ site system, in both cases the rank decomposition MPO construction algorithm produces an MPO of bond dimension $w = O(N)$. Unfortunately, due to its high complexity we were unable to use the algorithm to construct an MPO representation of $\widetilde{H}_\text{ks}$ for systems larger than $N = 9$. The bipartite graph construction algorithm Ref.~\cite{Ren-Li-Jiang-2020} is more efficient, but for $\hat{H}_\text{ks}$ it produces an MPO of bond dimension $w = O(N^2)$, which would severely limit our DMRG calculations.

Luckily, the bipartite graph algorithm can be modified to use a rank decomposition instead of the minimum vertex cover to combine the attractive features of both approaches. It retains the optimal bond dimensions of the rank-decomposition approach while also limiting $N_A$ in Eq.~\eqref{eq:two site Hamiltonian} to be proportional to the incoming bond dimension, which as previously mentioned is $O(N)$.

\begin{figure}
  \includegraphics[scale=0.6]{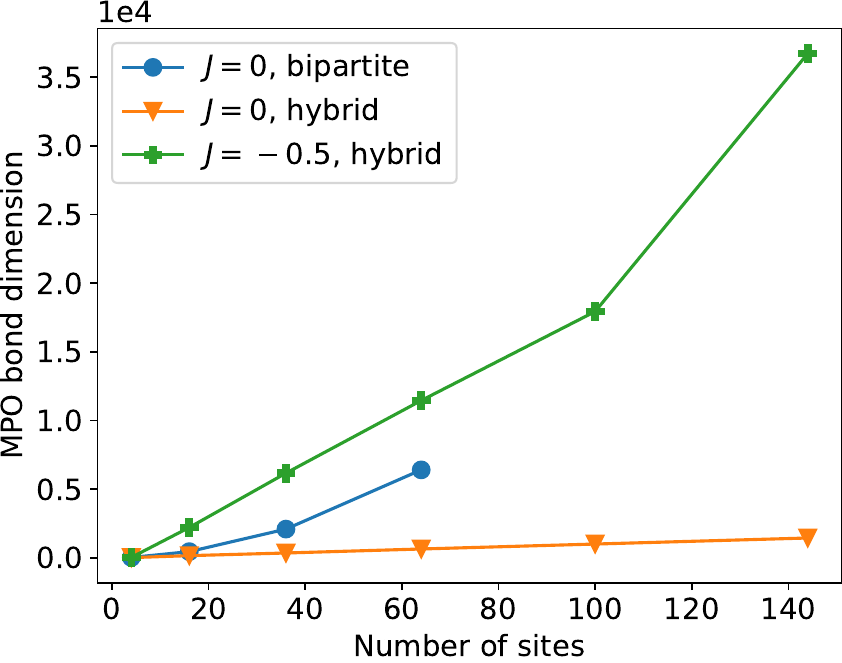}
  \caption{\label{fig:mpo-bond-dims} The MPO bond dimension of the (transcorrelated) momentum space Fermi-Hubbard Hamiltonian $\hat{H}_\text{ks} (\widetilde{H}_\text{ks})$ for square periodic lattices. MPOs were constructed with either the bipartite algorithm from `Renormalizer' \cite{Ren-Shuai-2018} or the hybrid algorithm from `ITensorMPOConstruction' \cite{ITensorMPOConstruction}. We discuss the anomaly for the $12 \times 12$ transcorrelated system in Appendix \ref{sec:Transcorrelated bond dimension}.}
\end{figure}

We believe this hybrid algorithm was first implemented in the `Block2' library \cite{Zhai-Larsson-Lee-2023}, we were unaware of it at the time and instead wrote our own open-source implementation in `ITensorMPOConstruction' \cite{ITensorMPOConstruction} that creates MPOs to be used with the `ITensor' library \cite{ITensors}. In Fig.~\ref{fig:mpo-bond-dims} we plot the bond dimension of $\hat{H}_\text{ks}$ for a square lattice with the bipartite graph algorithm and our hybrid algorithm, illustrating the behavior described above with the bipartite algorithm producing an MPO with a quadratic bond dimension, and the hybrid algorithm producing an MPO with a linear bond dimension. We also applied our hybrid algorithm to $\widetilde{H}_\text{ks}$ and again see the bond dimension grows linearly with the system size, albeit roughly 20 times faster than the bond dimension of $\hat{H}_\text{ks}$. The $12 \times 12$ transcorrelated system does deviate from the linear trend, but we argue in Appendix \ref{sec:Transcorrelated bond dimension} that this increase is due to numerical noise.

In addition to many optimizations that allowed us to reach system sizes of up to $12 \times 12$, including using the sparse rank-revealing QR decomposition \cite{Davis-2011}, we developed a novel approach such that the resulting MPO tensors have the optimal block diagonal sparsity. Constructing a block diagonal MPO is necessary when using a symmetry adapted MPS to enforce global $U(1)$ symmetries \cite{Sukhwinder-Pfeifer-Vidal-2011}, but if the MPO is very sparse it can also greatly speed up DMRG calculations. Previous approaches to constructing block sparse MPOs used the symmetries to bring $\gamma$ in Eq.~\eqref{eq:two site Hamiltonian} into block diagonal form. The rank decomposition is then done individually on each symmetry sector (block) to ensure that the resulting MPO tensors do not mix the symmetry sectors. However, by finding the connected components of the bipartite graph representation, we can bring $\gamma$ into block diagonal form with the maximum number of possible blocks. Not only can this result in a sparser MPO than using symmetries alone, but it does not require the user to input or even know about the symmetries of the underlying Hamiltonian. Our algorithm produces MPOs of equal sparsity for both the transcorrelated and non-transcorrelated momentum space Hamiltonians, with the sparsity increasing from 99.82\% for the $6 \times 6$ system to 99.96\% for the $12 \times 12$ system \footnote{
  Our connected components approach is guaranteed to produce the sparsest block diagonal form. However, many libraries also store the block diagonal blocks in a sparse format, and the QR decomposition we use to generate these blocks is not guaranteed to produce the greatest overall MPO sparsity. We have found that in cases where the rank decomposition algorithm does not result in a significantly smaller bond dimension than the bipartite approach, the bipartite algorithm is preferred due to the greater overall sparsity. The electronic structure Hamiltonian is the prime example of such a situation. For more details see the `ITensorMPOConstruction' documentation \cite{ITensorMPOConstruction}.
}.

For the $10 \times 10$ and $12 \times 12$ systems, the MPO for $\widetilde{H}_\text{ks}(J)$ is prohibitively expensive to construct. Instead of constructing it independently for each value of $J$, we construct an MPO for the three electron terms that is independent of $J$ then, for each value of $J$, we form the MPO for the one and two electron terms and take the appropriate linear combination to form $\widetilde{H}_\text{ks}(J)$.

\subsection{Entanglement Structure} \label{sec:entanglement structure}

Due to the nature of the mixed canonical form, Eq.~\eqref{eq:mixed canonical form}, MPSs are an inherently one dimensional representation of a quantum state, and they excel at representing states with small correlation lengths in one dimension. However, a state that has a small correlation length in two dimensions will have a large correlation length when mapped to the one dimensional MPS, and therefore will require a large bond dimension to represent accurately. For a square $L \times L$ lattice, one obvious mapping is the row-major mapping in which the degree of freedom at lattice site $(n_x, n_y)$ is mapped to the MPS tensor at position $L n_y + n_x$. Applied to the real space Fermi-Hubbard Hamiltonian, the row-major mapping makes the effective correlation length increase by a factor of $L$. With uniform hopping amplitude this is a good mapping, but consider the case when hopping along the $y$ direction is preferred over the $x$ direction. In this case, the correlation length along the $y$-axis will be longer than along the $x$-axis and the column-major mapping will therefore result in a shorter effective correlation length and likely a more compact MPS representation.

For all-to-all Hamiltonians, such as the electronic structure or momentum space Fermi-Hubbard Hamiltonians, predicting which degrees of freedom are most strongly correlated is not as simple. Nevertheless, there is a long history of using a naive mapping to calculate the mutual information matrix, which has been show to be insensitive to both the mapping used and the MPS bond dimension \cite{Rissler-Noack-White-2006}, and then optimizing a heuristic cost function based on the mutual information to obtain a new mapping with a hopefully smaller effective correlation length \cite{Legeza-Solyom-2003,Rissler-Noack-White-2006,Barcza-Legeza-2011}. The authors of Ref.~\cite{Baiardi-Reiher-2020} had success using the Fiedler optimization method \cite{Fiedler-Miroslav-1973} to produce a new mapping for the momentum space Fermi-Hubbard Hamiltonian. 

There are a couple challenges to this approach. First off, the optimization problem is a combinatorial one, and therefore impossible to solve in generality. Next, the cost function is a heuristic and there is no direct relationship between the cost of a mapping and the accuracy of DMRG with said mapping. Additionally, the calculated mapping can vary drastically as the parameters of the Hamiltonian (e.g. interaction strength) or the target symmetry sector (e.g. the number of electrons) are changed, prohibiting straightforward comparisons between systems. Finally, as demonstrated above MPO construction can be very expensive and would need to be repeated for each new mapping. The entanglement structure of the ground state of the momentum space Fermi-Hubbard Hamiltonian was investigated in Ref.~\cite{Ehlers-Solyom-Legeza-2015}, however they still optimized the heuristic cost function to find a good mapping. Here, we take it one step further and developed two new mappings, one for dilute systems and one at half-filling. These mappings address the concerns listed above and outperform the Fiedler mapping in terms of the energies obtained with DMRG.

For two-dimensional systems the values of $\epsilon(\V{k})$ come in shells, for example there are $2 L - 2$ momentum modes with $\epsilon(\V{k}) = 0$ in an $L \times L$ periodic square lattice. One implication of this shell structure is that for $U = 0$ the ground state of open-shell systems are highly degenerate. At half-filling all the modes with $\epsilon(\V{k}) > 0$ will be doubly occupied, but any distribution of the remaining $2 L - 2$ electrons in the $\epsilon(\V{k}) = 0$ shell at the Fermi surface will be a ground state. The introduction of the interaction term breaks this degeneracy, but to first order in the interaction strength the ground state is well approximated by a specific state in the ground space of the hopping term. That is to say, every mode below the Fermi surface remains doubly occupied, and every mode above the Fermi surface remains empty, but strong entanglement may exist within the shell at the Fermi surface. As the interaction strength increases the most prominent excitations will involve modes near the Fermi surface. Based on this, we propose the so called $\epsilon$ mapping in which we map the momentum modes to MPS tensors based on the decreasing value of $\epsilon(\V{k})$, such that the correlations within a shell and between neighboring shells remain ``local'' in the MPS.
\begin{figure}
  \includegraphics[scale=0.6]{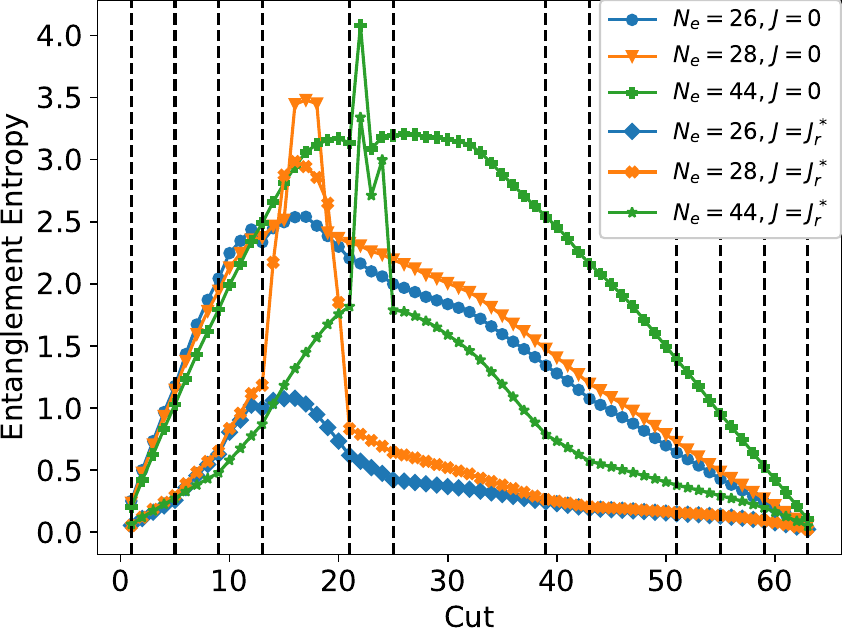}
  \caption{\label{fig:8x8-dilute-svn} The bipartite entanglement entropy with the $\epsilon$ mapping. Displayed is the entanglement entropy of each bipartition of the MPS approximation at $m = 1024$ to the ground state of the transcorrelated Hamiltonian for the $8 \times 8$ system with $N_e \in \{ 26, 28, 44 \}$ and $J \in \{ 0, J^*_r \}$, where $J^*_r$ is given in Table~\ref{tab:j values}. The vertical lines correspond to the shell structure of $\epsilon(\V{k})$. }
\end{figure}

In Fig.~\ref{fig:8x8-dilute-svn} we plot the bipartite entanglement entropy, Eq.~\eqref{eq:entanglement entropy definition}, for the ground states of the $8 \times 8$ system with $N_e \in \{ 26, 28, 44 \}$ as a function of the bond cut, for MPSs of bond dimension $m = 1024$ using the $\epsilon$ mapping. The entropy at the $j$-th cut is the entropy between the momentum modes mapped to the first $j$ MPS tensors and the remaining modes. We also superimpose vertical lines that separate the shell structure of $\epsilon(\V{k})$. For example, $N_e = 26$ is a closed-shell calculation with the thirteen momentum modes with $\epsilon(\V{k}) \geq 2$ doubly occupied in the free electron solution. Hence, there is a vertical line at the thirteenth cut. By contrast, $N_e = 28$ and $N_e = 44$ are open-shell calculations, with each having two electrons in the shell at the Fermi surface at $\epsilon(\V{k}) = \sqrt{2}$ and $\epsilon(\V{k}) = 2 - \sqrt{2}$ respectively. The two open-shell calculations have a sharp peak in the entanglement entropy for cuts within the Fermi surface, demonstrating the strength of the correlations within this shell. The closed-shell calculation with $N_e = 26$ shares a similar entanglement profile with the open-shell $N_e = 28$ calculation, but lacks the sharp peak within the Fermi surface. Returning to the perturbative argument outlined above, closed-shell systems have unique ground states and since the shell at the Fermi surface is fully occupied it does not immediately become entangled with the introduction of a weak interaction. For all three calculations the entanglement decreases with the distance from the Fermi surface, agreeing with our prediction that the majority of the correlations will be near the Fermi surface.

We also plot the entanglement entropy of the transcorrelated ground states in Fig.~\ref{fig:8x8-dilute-svn} with the values of $J$ from Table~\ref{tab:j values}. The general features of the entanglement described above are unchanged by transcorrelation, but the overall amount of entanglement is reduced. While the transcorrelated method does a good job of suppressing the entanglement away from the Fermi surface, for open-shell systems the correlations within the shell at the Fermi surface remain strong. For both of the open-shell systems the transcorrelated method only slightly reduces the maximum entanglement. As such, transcorrelated DMRG seems particularly well suited to closed-shell systems that lack the strong correlations within the Fermi surface.

\begin{figure}
  \includegraphics[scale=0.58]{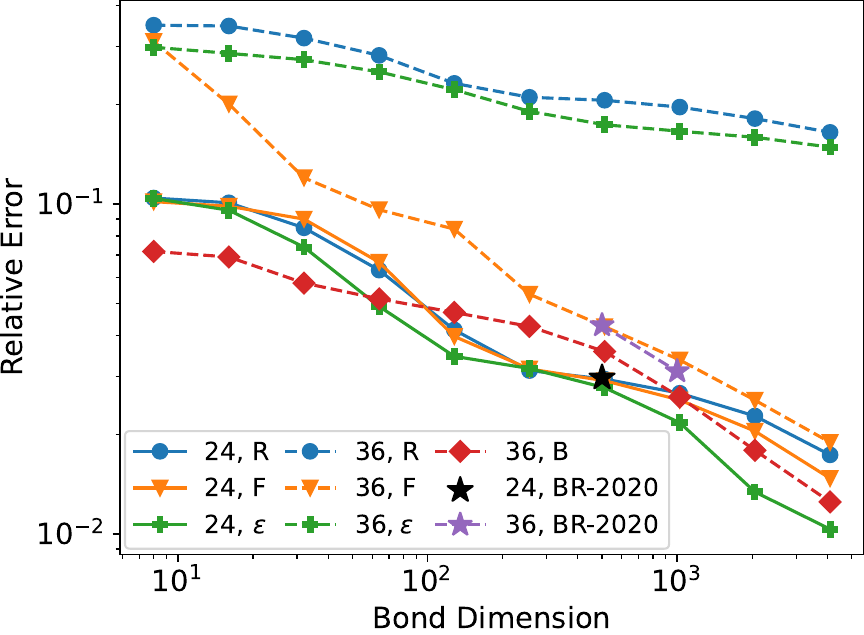}
  \caption{\label{fig:6x6-mapping-comparison} The relative error in the DMRG energy of the $6 \times 6$ system with $N_e \in \{ 24, 36 \}$ and the row-major (R), Fiedler (F) and $\epsilon$ mappings. For $N_e = 36$ we include the bipartite (B) mapping. We also show the three data points, one at $N_e = 24$ and two at $N_e = 36$, from Ref.~\cite{Baiardi-Reiher-2020} with the Fiedler mapping.}
\end{figure}
In Fig.~\ref{fig:6x6-mapping-comparison} we plot the relative error in the energy obtained with DMRG and various mappings as a function of the MPS bond dimension for the $6 \times 6$ system with $N_e = 24(36)$, which corresponds to one-third(half) filling. Both of these are open-shell systems. The relative error is calculated as
\begin{equation} \label{eq:relative energy error}
  \delta = \frac{E_\text{ref} - E}{E_\text{ref}} \ ,
\end{equation}
where $E_\text{ref}$ is the AFQMC reference energy reported in Table~\ref{tab:energies} and $E$ is the energy we obtain with DMRG. With $N_e = 24$ both row-major and Fiedler mapping perform well, with around a 1.5\% error at $m = 4096$, but the $\epsilon$ mapping does slightly better obtaining a 1\% error. With $N_e = 36$ the situation is different, the $\epsilon$ mapping marginally outperforms the row-major mapping, but neither is able to achieve an error below 14\%, the Fiedler mapping on the other hand achieves an error of 2\%. This suggests that the entanglement structure at half-filling differs from the dilute systems, and we need a different mapping to account for these new correlations.

At half-filling, in addition to the strong correlations within shell at the Fermi surface (for which $\epsilon(\V{k}) = 0$), we observed strong correlations between the momentum modes $\V{k}$ and $\V{k} + \V{\pi}$. This is notable because $\epsilon(\V{k}) = - \epsilon(\V{k} + \V{\pi})$, so these correlations become long range in the $\epsilon$ mapping, explaining its poor performance in this regime. These new correlations have a nice explanation; the lattices we consider here are bipartite lattices, and the real space representation of
\begin{equation} \label{eq:bipartite k pi combo}
  \left( \hat{c}^\dagger_{\V{k}, \uparrow} + \hat{c}^\dagger_{\V{k} + \V{\pi}, \uparrow} \right) \left( \hat{c}^\dagger_{\V{k}, \downarrow} - \hat{c}^\dagger_{\V{k} + \V{\pi}, \downarrow} \right)
\end{equation}
creates a spin-$\uparrow$ electron on one sublattice and a spin-$\downarrow$ electron on the other sublattice, such that no site is doubly occupied. We empirically observed that these correlations between $\V{k}$ and $\V{k} + \V{\pi}$ persist at half-filling even for small $U / t$. We hypothesize that these correlations are important for any system on a bipartite lattice that is near half-filling or has a large value of $U / t$. Our justification for why these correlations are not important for the dilute systems examined above is that the effective interaction strength is small. Given a product state in momentum space, $\ket{s}$, the energy is
\begin{equation}
  \braket{s | \hat{H} | s} = \braket{s | \hat{T} | s} + \frac{U}{N} N_\uparrow N_\downarrow \ ,
\end{equation}
which is to say the energy penalty of a free electron solution is proportional to not only the interaction strength $U$ but also quadratically on the filling factor $\eta$. Our intuition is backed by the fact that for these dilute systems if we increase the interaction strength we observe the correlations between $\V{k}$ and $\V{k} + \V{\pi}$ becoming more and more important.

Using these correlations we created the so-called bipartite mapping, in which we map the momentum modes $\V{k}$ and $\V{k} + \V{\pi}$ to adjacent MPS tensors and then arrange the pairs by increasing $|\epsilon(\V{k})|$. As shown in Fig.~\ref{fig:6x6-mapping-comparison}, at half-filling the bipartite mapping outperforms all other mappings. It is particularly accurate at small bond dimensions, where it achieves half the error of the Fiedler mapping. This surprising accuracy at small bond dimensions can be attributed to the fact that with the bipartite mapping, using Eq.~\eqref{eq:bipartite k pi combo} a ground state of the interaction term can be constructed as MPS of bond dimension 4, and when restricted to the $\V{K} = 0$ sector as an MPS of bond dimension 8. 

\begin{figure}
  \includegraphics[scale=0.6]{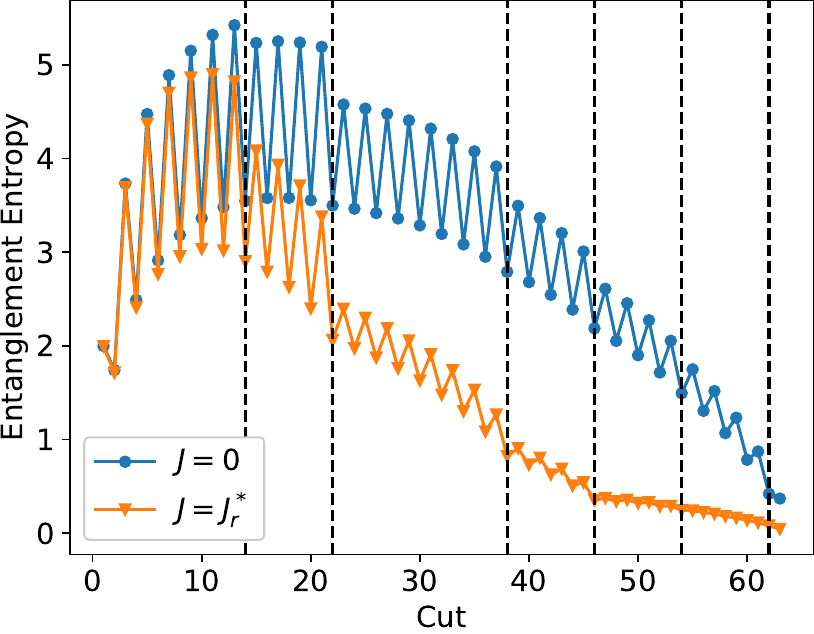}
  \caption{\label{fig:8x8-half-svn} The bipartite entanglement entropy with the bipartite mapping. Displayed is the entanglement entropy for each bipartition of the MPS approximation at $m = 512$ to the ground state of the transcorrelated Hamiltonian for the $8 \times 8$ system with $N_e = 64$ and $J \in \{ 0, J^*_r \}$, where $J^*_r$ is given in Table~\ref{tab:j values}. The vertical lines correspond to the shell structure of $|\epsilon(\V{k})|$. }
\end{figure}
In Fig.~\ref{fig:8x8-half-svn} we plot the bipartite entanglement entropy for the $8 \times 8$ system at half-filling using the bipartite mapping, along with vertical lines which denote the shell structure of $|\epsilon(\V{k})|$. The most obvious feature of the entanglement is the saw like appearance, with a larger entanglement for the odd cuts than the even cuts. This is due to the large entanglement between the momentum modes $\V{k}$ and $\V{k} + \V{\pi}$ which are mapped to adjacent MPS tensors; for the odd cuts one pair of $\V{k}$ and $\V{k} + \V{\pi}$ are split apart, leading to a much larger entanglement than the even cuts where $\V{k}$ and $\V{k} + \V{\pi}$ are always on one side or the other. The largest entanglement occurs in the $\epsilon(\V{k}) = 0$ shell at the beginning of the MPS, and the entanglement decreases as $|\epsilon(\V{k})|$ increases. Additionally, the entanglement between $\V{k}$ and $\V{k} + \V{\pi}$, illustrated by the magnitude of the oscillations, decreases as $|\epsilon(\V{k})|$ increases. In fact, the magnitude of these oscillations appear to be roughly constant within a shell, and it is possible to infer the shell structure of $|\epsilon(\V{k})|$. Compared to the bipartite entanglement of the $N_e = 44$ calculation shown in Fig.~\ref{fig:8x8-dilute-svn}, with $N_e = 64$ the maximum entanglement is larger and since the entanglement is not as sharply peaked, high entanglement persists across more cuts, lending further credence to the trend observed earlier that dilute ground states are more amenable to MPS approximation.

We also plot in Fig.~\ref{fig:8x8-half-svn} the bipartite entanglement entropy of the transcorrelated ground state of the $8 \times 8$ system at half-filling with the value of $J$ taken from Table~\ref{tab:j values}. As with the dilute systems, the transcorrelated method does a good job of reducing the entanglement for cuts far from the Fermi surface, but the entanglement for cuts within the shell at the Fermi surface remain strong. This trend also applies to the entanglement between $\V{k}$ and $\V{k} + \V{\pi}$; transcorrelation does little to reduce this entanglement when $\epsilon(\V{k}) = 0$, but as $|\epsilon(\V{k})|$ increases this entanglement is reduced, with the oscillations in the bipartite entanglement almost completely disappearing for the three shells with the largest values of $|\epsilon(\V{k})|$.

Looking back at Fig.~\ref{fig:6x6-mapping-comparison} we include the non-transcorrelated DMRG energies obtained for the $6 \times 6$ systems from Ref.~\cite{Baiardi-Reiher-2020} using the Fiedler mapping. For $N_e = 24$ the authors present energies obtained with $m = 500$, and using the $\epsilon$ mapping we obtain a lower energy at $m = 512$. For $N_e = 36$ the authors present energies obtained with $m \in \{500, 1000\}$ and again using the bipartite mapping we obtain lower energies at similar bond dimensions. Since these are non-transcorrelated calculations they are variational, and in that sense the lower energies we obtain truly are better.

For all remaining calculations, we use the $\epsilon$ mapping for dilute systems and the bipartite mapping at half-filling.

\subsection{Choice of J}

\begin{figure}
  \includegraphics[scale=0.57]{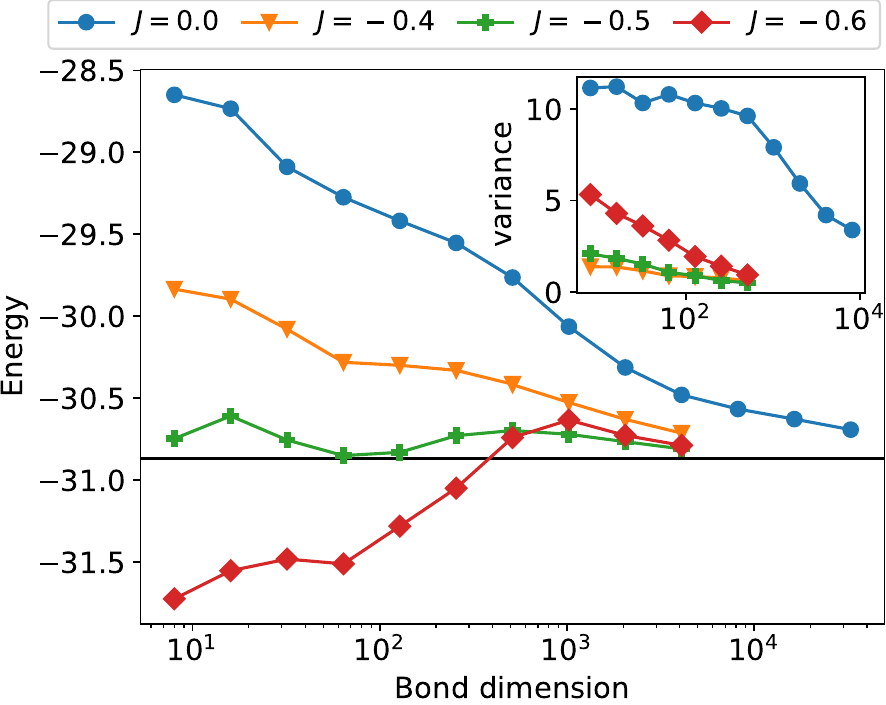}
  \caption{\label{fig:6x6-energy-sweep} The energy and variance obtained for the $6 \times 6$ transcorrelated system with $N_e = 36$ and $J \in \{ 0, -0.4, -0.5, -0.6 \}$ as a function of the MPS bond dimension. The horizontal line is the AFQMC reference energy given in Table~\ref{tab:energies}, with a width proportional to the reported uncertainty.}
\end{figure}

In Fig.~\ref{fig:6x6-energy-sweep} we plot the energy obtained for the $6 \times 6$ transcorrelated system at half-filling for $J \in \{ 0, -0.4, -0.5, -0.6 \}$ as a function of the MPS bond dimension. From this plot we can observe two important effects of the non-variational nature of the transcorrelated method; the transcorrelated energy is neither bounded from below by the ground state energy, nor does it converge monotonically with the bond dimension. These combine to make direct energy comparisons between calculations more difficult. For example, we find it unlikely that the approximation to the transcorrelated ground state with $J = -0.5$ is better at a bond dimension of 64 than at 4096, yet the former results in a more accurate ground state energy. A quantity that permits more direct comparison is the variance
\begin{equation} \label{eq:variance of j}
  \sigma^2(J) = \braket{\phi | \widetilde{H}(J)^\dagger \widetilde{H}(J) | \phi} - \left| \braket{\phi | \widetilde{H}(J) | \phi} \right|^2 ,
\end{equation}
which is variational and bounded from below by zero. We plot this in the inset to Fig.~\ref{fig:6x6-energy-sweep}. Aside from the anomaly at the start of the $J = 0$ calculation, the variance converges monotonically with the bond dimension for all values of $J$. This suggests that the non-monotonic behavior of the energy is inherent to these transcorrelated calculations, and not simply an artifact of our DMRG procedure having trouble optimizing the approximation to the ground state. Additionally, all the transcorrelated calculations have a much smaller variance than the $J = 0$ calculation, with the $J = -0.4$ and $J = -0.5$ calculations at $m = 8$ producing a smaller variance than the $J = 0$ calculation at $m = 32768$.

While we reached a bond dimension of 32768 for $J = 0$, we were only able to go up to a bond dimension of 4096 for the transcorrelated calculations because of the almost $20 \times$ increase in the MPO bond dimension. Running on a system with two Intel Xeon E7-4809 v3 and 3TB of RAM, the $J = 0$ calculation at a bond dimension of 32768 took 16 hours per sweep of DMRG, whereas a transcorrelated calculation at $m = 4096$ took 15 hours per sweep. For $J \in \{-0.4, -0.5, -0.6\}$, a bond dimension $m$ transcorrelated calculation always produces a more accurate energy than the $J = 0$ energy at the same bond dimension, but they are not always more accurate than the similarly expensive $J = 0$ energy at a bond dimension of $8 m$. On the other hand, not until a bond dimension of 32768 does the $J = 0$ calculation exceed the accuracy of the $J = -0.5$ transcorrelated calculation \emph{at any bond dimension}.

This elucidates the need for a procedure to choose a ``good'' value of $J$ in order to maximize the utility of the transcorrelated method. Furthermore, this procedure needs to work without knowledge of any reference energy. We consider two such criteria to set the value of $J$ \footnote{
  A third method not used here, can also be used to optimize the correlator. This is done by minimizing the energy
  \begin{equation*}
    E_\text{VMC}(J) = \frac{\braket{\phi | e^{J \hat{D}} \hat{H} e^{J \hat{D}} | \phi}}{\braket{\phi | e^{2 J \hat{D}} | \phi}} ,
  \end{equation*}
  with variational Monte Carlo (VMC). Though on its own VMC is distinct from the transcorrelated method, it has been combined with the transcorrelated self-consistent field equations to optimize $\ket{\phi}$ \cite{Umezawa-Tsuneyuki-2003, Ochi-2023}.
}.
The first is to choose the value that minimizes the variance, Eq.~\eqref{eq:variance of j}, which was used in Refs.~\cite{Tsuneyuki-2008,Haupt-Hosseini-Mohammadreza-2023,Liao-Zhai-Christlmaier-2023} and is variational due to its Hermitian nature. The second criterion dates back to the first transcorrelated papers by Boys and Handy, Ref.~\cite{Boys-Handy-1969c}, and is also used in Refs.~\cite{Luo-2010, Luo-2011, Wahlen-Strothman-Hoyos-2015, Neuscamman-Changlani-Kinder-2011,Dobrautz-Luo-Alavi-2019}. It picks $J$ such that the residual
\begin{equation} \label{eq:projected residual}
  r(J) = \braket{\phi | \hat{D} \widetilde{H}(J) | \phi} - \braket{\phi | \hat{D} | \phi} \braket{\phi | \widetilde{H}(J) | \phi} ,
\end{equation}
is zero, and lacks a variational guarantee. With one exception, all prior works optimize the correlator against a single Slater determinant reference state. The most sophisticated approach, taken by Luo in Refs.~\cite{Luo-2010, Luo-2011}, alternatively optimizes both the correlator using Eq.~\eqref{eq:projected residual} and the reference state $\ket{\phi}$ which is represented as a linear combination of up to 660 Slater determinants. Their state optimization was done using complete active space self-consistent field theory and required a Hermitian truncation of the transcorrelated Hamiltonian. Inspired by their success, we let $\ket{\phi}$ be the MPS approximation to the ground state of the full transcorrelated Hamiltonian. In this manner the optimization is self-consistent, and is redone when we increase the bond dimension of $\ket{\phi}$.

\begin{figure}
  \includegraphics[scale=0.57]{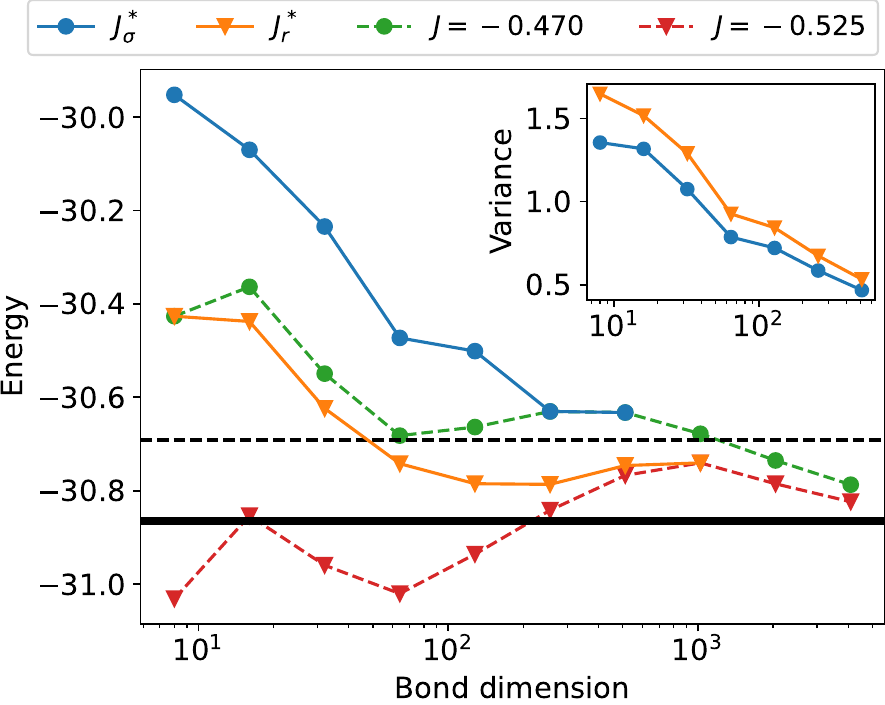}
  \caption{\label{fig:6x6-j-opt-energy} The energy obtained for the $6 \times 6$ transcorrelated system with $N_e = 36$ and $J$ obtained by minimizing the variance, $J^*_\sigma$, and by finding the zero of the residual, $J^*_r$, at each bond dimension. These values of $J^*$ and their corresponding energies are given in Tables~\ref{tab:j values} and \ref{tab:energies}. Also shown are calculations with fixed $J = -0.470(-0.525)$ which is the optimal value $J^*_\sigma(J^*_r)$ calculated at $m = 256(1024)$. The solid horizontal line is the AFQMC reference energy given in Table~\ref{tab:energies}, with a width proportional to the reported uncertainty. The dashed horizontal line is the energy from the $J = 0$ calculation at $m = 32768$. The insert shows the variance obtained with both $J^*_\sigma$ and $J^*_r$.}
\end{figure}

We found that the variance, Eq.~\eqref{eq:variance of j}, is roughly quadratic while the residual, Eq.~\eqref{eq:projected residual}, is roughly linear in $J$ regardless of the bond dimension of $\ket{\phi}$. In Fig.~\ref{fig:6x6-j-opt-energy} we plot the energies $E(J^*_\sigma)$ and $E(J^*_r)$ obtained by optimizing the variance and residual respectively at each bond dimension for the $6 \times 6$ system at half-filling. Most importantly, both energies never dip below the ground state energy. In the insert to Fig.~\ref{fig:6x6-j-opt-energy}, we plot the variances $\sigma^2(J^*_\sigma)$ and $\sigma^2(J^*_r)$, both of which decrease monotonically with the bond dimension. In particular even though $E(J^*_\sigma)$ remains unchanged and $E(J^*_r)$ actually increases from bond dimension $m = 256$ to $m = 512$, both variances continue to decrease.

The solid horizontal line in Fig.~\ref{fig:6x6-j-opt-energy} is the AFQMC reference energy given in Table~\ref{tab:energies} with a width proportional to the uncertainty, and the dashed horizontal line is the energy from the $J = 0$ calculation at $m = 32768$. Though we were unable to obtain energies that converged to within the AFQMC error bars, we exceeded the accuracy of this $J = 0$ calculation with $J^*_r$ at $m = 64$. Additionally, we plot the energy obtained when we fix $J = -0.470$ which is $J^*_\sigma$ calculated at $m = 512$ and $J = -0.525$ which is $J^*_r$ calculated at $m = 1024$. In both cases, using the optimal value $J^*$ at each bond dimension results in smoother energy curves than using a fixed, albeit ``good'', value of $J$. However, by fixing $J$ we were able to reach larger bond dimensions. For both of these fixed values, by $m = 1024$ we see what appears to be smooth monotonic convergence to the ground state energy, and by $m = 2048$, we exceed the accuracy of the $J = 0$ calculation at $m = 32768$. Our most accurate transcorrelated calculation, with $J = -0.525$ at $m = 4096$, has a relative energy error, Eq.~\eqref{eq:relative energy error}, of 0.13\%, whereas the similarly expensive $J = 0$ calculation at $m = 32768$ had an error of 0.56\%, an improvement by over $4 \times$.

\begin{figure}
  \includegraphics[scale=0.55]{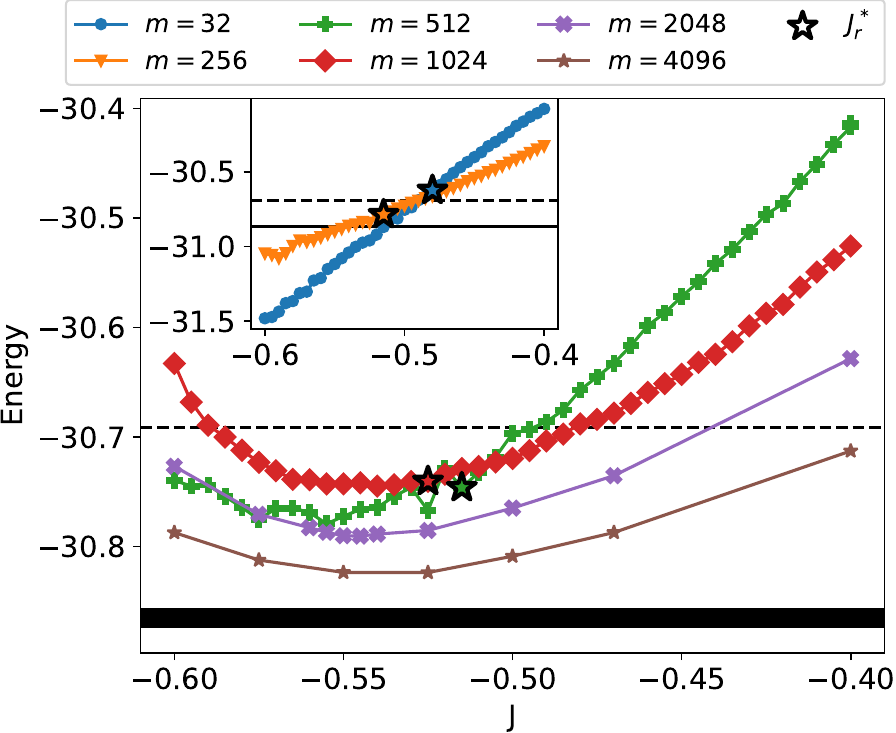}
  \caption{\label{fig:6x6-energy-vs-j} The transcorrelated energy obtained with DMRG as function of $J$ for the $6 \times 6$ system with $N_e = 36$ at various bond dimensions $m$. We also plot the optimal values of $J^*_r$ obtained by optimizing the residual. These values are given in Table~\ref{tab:j values}. The horizontal lines are the AFQMC reference energy given in Table~\ref{tab:energies}, with a width proportional to the reported uncertainty. The dashed horizontal lines are the energy from the $J = 0$ calculation at $m = 32768$.}
\end{figure}

Using Eq.~\eqref{eq:transcorrelated hopping term} and the fact that $\hat{T}_{sd} = \hat{T}_{ds}^\dagger$ we have
\begin{equation} \label{eq:energy as a function of j}
  \Re \left[ \braket{\phi | \left( \widetilde{H}(J) - \hat{H} \right) | \phi} \right] \propto \cosh(J) - 1 \ ,
\end{equation}
which means that for fixed $\ket{\phi}$ the expectation value of the transcorrelated Hamiltonian is unbounded in $J$. This implies that for any $\ket{\phi}$ for which the constant of proportionality in Eq.~\eqref{eq:energy as a function of j} is negative (as is the case for all our MPS ground state approximations), it is possible to find a value of $J$ such that the transcorrelated energy, $\braket{\phi | \widetilde{H}(J) | \phi}$, is equal to the exact ground state energy. In Fig.~\ref{fig:6x6-energy-vs-j} we plot the transcorrelated energy as a function of $J$, but instead of keeping $\ket{\phi}$ fixed, it is the DMRG approximation to the ground state of $\widetilde{H}(J)$. At small bond dimensions we observed the expected behavior where the transcorrelated energy appears to have no minima. However, at $m \ge 512$ the energy is bounded from below by the ground state energy. Additionally, the slope of the energy curves is decreasing with the bond dimension, which is expected since in the large bond dimension limit the ground state of $\widetilde{H}(J)$ can be represented exactly regardless of $J$. Starting at a bond dimension of $1024$ the energy appears to converge smoothly at all values of $J$ as the bond dimension increases.

We also plot the optimal values of $J^*_r$ in Fig.~\ref{fig:6x6-energy-vs-j}, and we see that at $m = 1024$ this optimal value is near the minimum in the energy, this leads to two observations. First, at $m = 1024$, no other value of $J$ yields a significantly more accurate energy. And second, perhaps at large bond dimensions the transcorrelated energy could be optimized directly, bypassing the need for the evaluation of the residual. Unfortunately, already for the $8 \times 8$ system at half-filling we only observe linear energy curves $E(J)$, presumably because are not able to run calculations at large enough bond dimensions. 

For the remaining systems, we choose to only optimize the residual. While optimizing the variance results in a smoother, potentially monotonic, energy curve, the residual optimization produces a more accurate energy and is more efficient to compute. Additionally, the variance $\sigma^2(J^*_r)$ is only slightly larger than $\sigma^2(J^*_\sigma)$. 

\section{Large-scale results} \label{sec:Results}

Our large-scale transcorrelated calculations are summarized in Tables~\ref{tab:j values} and \ref{tab:energies}. In Table~\ref{tab:j values} we present the optimal values $J^*$ calculated at each bond dimension for the systems considered in this paper. Only for the $6 \times 6$ system at half-filling did we optimize both the variance and residual. Our optimization procedure was simple; we calculated each function at intervals of $\Delta J = 0.005$. We found this spacing small enough to localize the optimal value such that the variation in energy between consecutive values of $J$ was small. Table~\ref{tab:energies} contains the energies obtained at each bond dimension $m$ using the value of $J^*_r$ from Table~\ref{tab:j values}. The reference energies calculated with AFQMC \cite{Qin-Shi-Zhang-2016,Shi-Zhang-2013} along with the reported statistical uncertainties are given as well. At half-filling (the upper section of the table) the AFQMC calculations are sign problem free, but off-half-filling (the lower section of the table) there is an additional bias that is not accounted for in the reported uncertainty. For all systems, due to the optimization of $J$, the energy we obtain with DMRG never drops below the reference energy. However, for both the $6 \times 6$ and $8 \times 8$ systems at half-filling we notice the energy increases at $m = 512$; we postulate that the larger half-filled systems would also exhibit non-monotonic behavior if were able to reach larger bond dimensions. Fortunately, for the dilute systems the non-monotonic behavior only appears for the $8 \times 8$ system with $N_e = 44$, and it is greatly reduced.
\begin{table*}
  \centering
  \begin{threeparttable}
    \caption{\label{tab:j values} The optimal values $J^*$ calculated at each bond dimension $m$ for square lattices with periodic boundary conditions and $U / t = 4$ in the $S_z = \V{K} = 0$ sector. The variance, Eq.~\eqref{eq:variance of j}, is only optimized for the $6 \times 6$ system with $N_e = 36$. The residual, Eq.~\eqref{eq:projected residual}, is optimized for all systems. The optimal values were determined up to $\Delta J = 0.005$.}
    \begin{tabular}{llllllllllll}
      \toprule
      System         & $N_e$ & Criteria      & $m = 8$ & $m = 16$ & $m = 32$ & $m = 64$ & $m = 128$ & $m = 256$ & $m = 512$ & $m = 1024$ \\ \midrule
      $6 \times 6$   & 36    & $\sigma^2(J)$ & -0.415  & -0.430   & -0.425   & -0.435   &  -0.440   & -0.470    & -0.470    &            \\ \midrule
      $6 \times 6$   & 36    & $r(J)$        & -0.470  & -0.475   & -0.480   & -0.480   &  -0.490   & -0.515    & -0.515    & -0.525     \\ \midrule
      $8 \times 8$   & 64    & $r(J)$        & -0.470  & -0.470   & -0.470   & -0.475   & -0.485    & -0.505    & -0.515    &            \\ \midrule
      $10 \times 10$ & 100   & $r(J)$        & -0.495  & -0.475   & -0.470   & -0.470   & -0.480    & -0.495    &           &            \\ \midrule
      $12 \times 12$ & 144   & $r(J)$        & -0.515  & -0.475   & -0.470   & -0.470   & -0.475    &           &           &            \\ \midrule \midrule
      $6 \times 6$   & 24    & $r(J)$        & -0.520   & -0.525  & -0.535   & -0.540   & -0.555    & -0.545    & -0.540    & -0.535     \\ \midrule
      $8 \times 8$   & 26    & $r(J)$        & -0.470  & -0.470   & -0.470   & -0.500   & -0.495    & -0.490    & -0.490    & -0.495     \\ \midrule
      $8 \times 8$   & 28    & $r(J)$        & -0.480  & -0.480   & -0.485   & -0.490   & -0.490    & -0.495    & -0.495    & -0.500     \\ \midrule
      $8 \times 8$   & 44    & $r(J)$        & -0.530  & -0.530   & -0.530   & -0.545   & -0.540    & -0.545    & -0.550    & -0.550     \\ \midrule
      \bottomrule
    \end{tabular}
  \end{threeparttable}
\end{table*}
\begin{table*}
  \centering
  \begin{threeparttable}
    \caption{\label{tab:energies} The ground state energies obtained with DMRG of the transcorrelated momentum space Fermi-Hubbard Hamiltonian on a square lattice using the value of $J^*_r$ calculated at each bond dimension $m$ (see Table~\ref{tab:j values}). We set $t = 1$, $U = 4$ and look in the $S_z = \V{K} = 0$ sector. The reference energies are calculated with AFQMC \cite{Qin-Shi-Zhang-2016,Shi-Zhang-2013}. Additionally, we calculate the relative error in the transcorrelated energy at the largest bond dimension, $\widetilde{\delta}$, as well as the transcorrelated variance, $\widetilde{\sigma}^2$, at one quarter of the largest bond dimension. We compare these with the relative error in the non-transcorrelated energy, $\delta$, and the non-transcorrelated variance, $\sigma^2$, both calculated at eight times the bond dimension of their transcorrelated counterparts.}
    \begin{tabular}{ccccccccccccccc}
      \toprule
                     & $N_e$ & $m = 8$ & $m = 16$ & $m = 32$ & $m = 64$ & $m = 128$ & $m = 256$ & $m = 512$ & $m = 1024$ & $E_\text{ref}$ & $\widetilde{\delta}$ (\%) & $\frac{\widetilde{\sigma}}{|E_\text{ref}|}$ (\%) & $\frac{\delta}{\widetilde{\delta}}$ & $\frac{\sigma}{\widetilde{\sigma}}$ \\ \midrule
      $6 \times 6$   & 36    & -30.43  & -30.44   & -30.62   & -30.74   & -30.79    & -30.79    & -30.75    & -30.74     & -30.865(9)     & 0.40                      & 2.66                                             & 2.40                                & 2.97                                \\ \midrule
      $8 \times 8$   & 64    & -54.29  & -54.30   & -54.46   & -54.59   & -54.67    & -54.81    & -54.74    &            & -55.05(1)      & 0.56                      & 2.67                                             & 4.85                                & 2.80                                \\ \midrule
      $10 \times 10$ & 100   & -85.02  & -84.92   & -85.00   & -85.14   & -85.33    & -85.55    &           &            & -86.12(4)      & 0.67                      & 2.46                                             & 6.95                                & 2.62                                \\ \midrule
      $12 \times 12$ & 144   & -122.10 & -122.27  & -122.36  & -122.56  & -122.72   &           &           &            & -123.95(2)     & 0.99                      & 2.18                                             & 6.35                                & 2.48                                \\ \midrule \midrule
      $6 \times 6$   & 24    & -42.01  & -42.07   & -42.09   & -42.34   & -42.52    & -42.57    & -42.60    & -42.63     & -42.669(1)     & 0.08                      & 2.21                                             & 10.76                               & 2.64                                \\ \midrule
      $8 \times 8^*$ & 26    & -66.47  & -66.50   & -66.51   & -66.73   & -66.82    & -66.82    & -66.83    & -66.84     & -66.855(2)     & 0.02                      & 0.93                                             & 14.24                               & 3.84                                \\ \midrule
      $8 \times 8$   & 28    & -68.08  & -68.09   & -68.15   & -68.20   & -68.37    & -68.49    & -68.51    & -68.55     & -68.595(1)     & 0.07                      & 1.33                                             & 6.97                                & 3.15                                \\ \midrule
      $8 \times 8$   & 44    & -74.95  & -74.98   & -74.94   & -74.91   & -75.11    & -75.55    & -75.67    & -75.72     & -75.89(1)      & 0.23                      & 2.24                                             & 8.84                                & 2.97                                \\ \midrule
      \bottomrule
    \end{tabular}
    \begin{tablenotes}              
      \small
      \item[*] Closed shell system
  \end{tablenotes}
  \end{threeparttable}
\end{table*}

At half-filling the relative error in the transcorrelated energy, $\widetilde{\delta}$, calculated using Eq.~\eqref{eq:relative energy error} and the transcorrelated energy obtained at the largest bond dimension, increases with the system size from 0.4\% for the $6 \times 6$ system up to almost 1\% for the $12 \times 12$ system. This behavior persists even if we calculate the relative error at the same bond dimension for each system, yielding the unsurprising conclusion that to hit a target relative error, larger systems with more electrons require a larger bond dimension. The energies are also far from converging to within the error bars of the AFQMC calculations, which would require a relative error of less than 0.05\% for the half-filled systems. In this regard the transcorrelated method again performs better when applied to  dilute systems, where the relative error is smaller by at least a factor of two compared to half-filling. 

In addition, we calculated the variance of the transcorrelated ground state estimate using Eq.~\eqref{eq:variance of j}, here we denote this quantity as $\widetilde{\sigma}^2$. Because this calculation is costly, if the maximum bond dimension we reached for a system was $m$, we calculated the variance at a bond dimension of $m / 4$. In order to better compare across different system sizes, in Table~\ref{tab:energies} we present the normalized standard deviation $\frac{\widetilde{\sigma}}{|E_\text{ref}|}$. This value is fairly constant at around 2.5\% for the half-filled systems, even though the relative error in the energy increases with system size. For the dilute systems, the normalized standard deviation is smaller than at half-filling, but not by as much as one might expect. Take the $8 \times 8$ system with $N_e = 26$, the relative energy error has decreased by a factor of 28 from the half-filled system, but the normalized standard deviation only decreased by a factor of $2.9$.

We also compare the transcorrelated errors and variances, $\widetilde{\delta}$ and $\widetilde{\sigma}^2$ with their non-transcorrelated ($J = 0$) counterparts, $\delta$ and $\sigma^2$, calculated at an eight times larger bond dimension. This constitutes a fair comparison since we found a transcorrelated calculation at a bond dimension of $m$ takes strictly less time than a non-transcorrelated calculation of bond dimension $8 m$ due to the almost 20 times increase in the bond dimension of the transcorrelated MPO. By using the transcorrelated method we obtained energies 2.4 to 14 times more accurate and standard deviations 2.5 to 3.8 times smaller than the non-transcorrelated calculations. A twofold improvement may not sound substantial, but due to the slow rate of convergence of the non-transcorrelated calculations it is significant. Consider the $8 \times 8$ system at half-filling. At $m = 512$ the relative energy error obtained with the non-transcorrelated method are $\delta = 5.4\%$, but by $m = 4096$ they have only decreased to $\delta = 2.7\%$. Assuming the non-transcorrelated error continues to decrease by a factor of two for every $8 \times$ increase in the bond dimension, the non-transcorrelated calculation won't match the transcorrelated accuracy until a bond dimension of $4096 \times 4.85^3 \approx 470,000$.

In Sec.~\ref{sec:entanglement structure} we demonstrated that the non-transcorrelated method is best suited for dilute closed-shell systems, and interestingly, the improvement obtained with the Gutzwiller ansatz is also largest for these same systems. The dilute systems all have the largest improvement in the relative error, and the closed-shell $8 \times 8$ system with $N_e = 26$ has the largest improvement in both relative error and standard deviation. While the normalized standard deviations we obtain, for both the transcorrelated and non-transcorrelated systems, are much larger than the relative errors in the energy, this is expected since the variance of MPS approximations has been shown to converge slower than the energy \cite{Silvester-Carleo-White-2025}.

Previous work on applying the transcorrelated method and DMRG to the Fermi Hubbard model, Ref.~\cite{Baiardi-Reiher-2020}, presented results for the $6 \times 6$ systems using the Fiedler mappings and the values of $J$ calculated in Ref.~\cite{Dobrautz-Luo-Alavi-2019} by optimizing the residual, Eq.~\eqref{eq:projected residual}, of a free electron solution. With $N_e = 24$ and an MPS of bond dimension $m = 500$, the authors obtained a transcorrelated relative error of $\widetilde{\delta} = 0.41\%$ using $J = -0.52372$. In contrast, we obtained $\widetilde{\delta} = 0.16\%$ at $m = 512$ and $J = -0.54$. We attribute our increased accuracy primarily to the self-consistent optimization of $J$, since as illustrated in Fig.~\ref{fig:6x6-mapping-comparison} the $\epsilon$ mapping provides only a modest improvement over the Fiedler mapping. With $N_e = 36$ the authors obtain an error of $\widetilde{\delta} = -0.26 (-0.07)\%$ at $m = 500 (1000)$. Despite the more substantial improvement from the bipartite mapping, at $m = 1024$ we obtained a much larger error of $\widetilde{\delta} = 0.4\%$. Note however that the energies the authors obtained are both below the ground state energy. While based on these two data points the energy appears to be converging to the ground state energy, it is conceivable that at larger bond dimensions the energy overshoots the ground state energy. For this calculation the authors used $J = -0.58521$, and as demonstrated in Figs.~\ref{fig:6x6-energy-sweep} and \ref{fig:6x6-j-opt-energy} we observed oscillations in the energy for $J \leq -0.47$, with these oscillations increasing as $J$ decreases. As shown in Fig.~\ref{fig:6x6-energy-vs-j} it is relatively straightforward to hand-pick a state whose transcorrelated energy is equal to a target ground state energy, so that direct comparisons between the energies of transcorrelated calculations may not be the whole story. The variance is also a relevant metric to compare since it is variational, but unfortunately this data was not available in Ref.~\cite{Baiardi-Reiher-2020}.

\section{Conclusion} \label{sec:Conclusion}

Our work on approximating the ground state of the two-dimensional momentum space Fermi Hubbard Hamiltonian as a Gutzwiller correlated MPS using the transcorrelated method is based on a combination of three technical innovations. First, we designed a state-of-the-art generic MPO construction algorithm that allowed us to construct MPOs for Hamiltonians consisting of tens of billions of terms. This enabled us to apply the transcorrelated method to larger lattices than previously possible. Second, inspired by the entanglement structure of the ground state we created two new mappings from momentum space modes to MPS tensors, the $\epsilon$ mapping for dilute systems and the bipartite mapping for half-filled systems. These new mappings not only make comparisons between different systems straightforward and illuminate the underlying physics, but they perform better than the competition when it comes to the accuracy of DMRG. Finally, we developed a framework to optimize the parameter of the Gutzwiller correlator alongside the MPS representation of the transcorrelated ground state. This optimization not only sped up the convergence to the ground state energy, but also unlike prior work we never obtain energies below the ground state energy. In addition, this optimization was only tractable because our MPO construction algorithm produced MPOs of optimal bond dimension and sparsity.

With these three innovations we were able to obtain a relative energy error of 0.02\% for the $8 \times 8$ system with 26 electrons, an improvement by a factor of 14 over a computationally equivalent non-transcorrelated calculation. However, this was not only the most dilute system we examined, but also a closed-shell system, features that both improve the performance of the transcorrelated method. For the open-shell systems, and in particular at half-filling, the transcorrelated method struggles at the fundamental challenge of reducing the entanglement within the shell at the Fermi surface, limiting the accuracy of the MPS approximation to the ground state. Nevertheless, in this regime we still demonstrated its benefit, obtaining energies that were 2.4 to 7.0 times more accurate than the equivalent non-transcorrelated energies.

Perhaps by refining the techniques used here, such as using a non-uniform bond dimension MPS, optimizing the DMRG algorithm for highly sparse MPOs, and using multi-node and/or GPU parallelism one could reach bond dimensions an order of magnitude larger than those obtained here \cite{Menczer-van-Damme-Rask-2024}. Or by further studying the entanglement structure of the ground state better mappings to MPS tensors could be crafted, and better initial states could be used to increase the accuracy at a given bond dimension. Nevertheless, we expect obtaining a relative energy error of $10^{-4}$ for even the $8 \times 8$ system will require novel techniques. Perhaps a more complicated correlator, one specifically designed to reduce the correlations within the shell at the Fermi surface would be more successful. In this regard, both our first and third innovations, should be directly applicable to any other transcorrelated calculation, with the optimal MPO construction algorithm and correlator optimization working in tandem to enable the application of the method to larger systems with more complicated and fine-tuned correlators.


\begin{acknowledgments}
This work was supported by the National Science Foundation CHE-2037832. Support is also acknowledged from the U.S. Department of Energy, Office of Science National Quantum Information Science Research Center, Quantum Systems Accelerator.

We would like to thank the UNM Center for Advanced Research Computing, supported in part by the National Science Foundation, for providing the high performance computing resources used in this work.
\end{acknowledgments}

\appendix

\section{DMRG procedure} \label{sec:DMRG procedure}

\begin{figure*}
  \centering
  \subfloat[]{
    \includegraphics[scale=0.55]{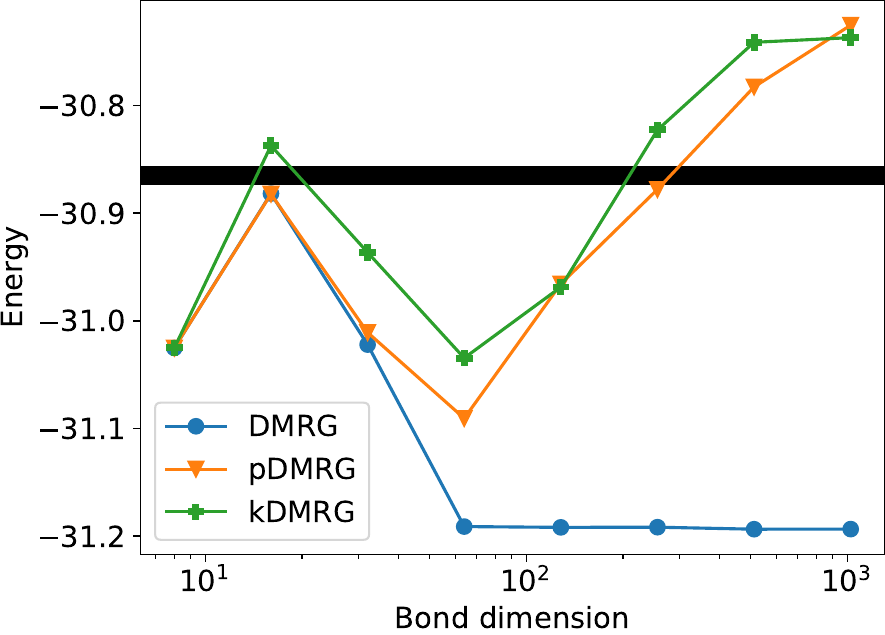}
    \label{fig:6x6-dmrg-comparison}
  }
  \subfloat[]{
    \includegraphics[scale=0.55]{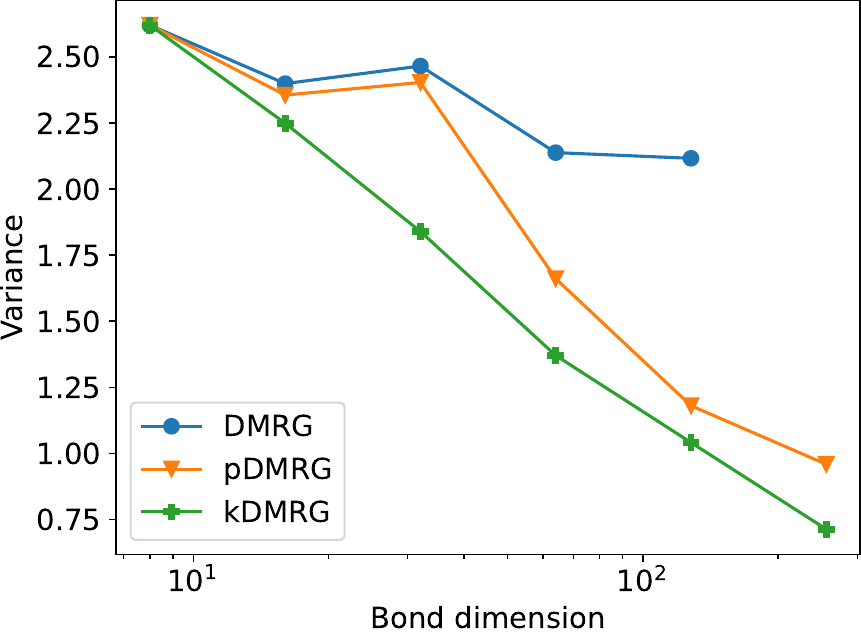}
    \label{fig:6x6-dmrg-comparison-variance}
  }
  \caption{\label{fig:6x6-dmrg-comparison-both} The (a) energy and (b) variance obtained for the $6 \times 6$ system with $N_e = 36$ and $J = -0.525$ with plain DMRG, perturbed DMRG (pDMRG) and Krylov DMRG (kDMRG). The horizontal line in (a) is the AFQMC reference energy given in Table~\ref{tab:energies}, with a width proportional to the reported uncertainty.}
\end{figure*}

For long range systems, such as the transcorrelated momentum space Fermi-Hubbard Hamiltonian, DMRG is prone to getting stuck in local minima. As shown in Fig.~\ref{fig:6x6-dmrg-comparison} for the $6 \times 6$ system at half-filling with $J = -0.525$, using eight sweeps of DMRG at each bond dimension results in DMRG getting stuck in a local minimum at a bond dimension of 64, which it is then unable to escape even as the bond dimension increases up to 1024. Alleviating this problem is an active area of research \cite{Hubig-McCulloch-2015, Gleis-Li-2023, McCulloch-Osborne-2024}, but a common approach is the density matrix perturbation introduced in Ref.~\cite{White-2005} and later adapted to two-site DMRG. Using four sweeps of perturbed DMRG (pDMRG) with an initial perturbation strength of $10^{-8}$ and decreasing by an order of magnitude each sweep followed by four sweeps of DMRG allows the calculation to escape the local minimum. The problem with pDMRG is that it is more expensive than DMRG; for this system and $m = 1024$, a single sweep with the perturbation takes five times longer than a sweep without. This overhead is clearly worth paying, but since the optimization of $J$ requires many runs of DMRG we could really benefit from a more efficient procedure.

Inspired by the success of Krylov subspace expansion for time evolution, Ref.~\cite{Yang-White-2020}, we tried diagonalizing the Hamiltonian in a two-dimensional Krylov subspace as the first step in an increased bond dimension calculation. Starting from an MPS approximation to the ground state of bond dimension $m$, $\ket{\psi_m}$, we form $\ket{r_{2m}} \approx (\hat{H} - \braket{\psi_m | \hat{H} | \psi_m}) \ket{\psi_m}$. Then we find the ground state of the Hamiltonian in the span of $\ket{\psi_m}$ and $\ket{r_{2m}}$ yielding $\ket{\psi'_{3m}}$. Finally, we run four sweeps of DMRG starting from $\ket{\psi'_{3m}}$ truncating back down to a bond dimension of $2m$ to yield $\ket{\psi_{2m}}$. As shown in Fig.~\ref{fig:6x6-dmrg-comparison}, this Krylov DMRG (kDMRG) procedure is also successful at avoiding the local minimum. To do the MPO-MPS contraction required to calculate $\ket{r_{2m}}$ efficiently we first use the zip-up algorithm to get a rough approximation followed by two sweeps of the variational algorithm for fine-tuning \cite{Paeckel-Kohler-Swoboda-2019}. At $m = 1024$ the overhead from this procedure is equivalent to two sweeps of DMRG. Not only is kDMRG more efficient at large bond dimensions than pDMRG, but it is also more accurate. As we show in Fig.~\ref{fig:6x6-dmrg-comparison-variance} the variance obtained with kDMRG is consistently smaller than the variance obtained with either DMRG or pDMRG.

Previous work on approximating the ground state of the transcorrelated Fermi-Hubbard Hamiltonian as an MPS, Ref.~\cite{Baiardi-Reiher-2020}, used imaginary time evolution approximated through the time dependent variational principle (TDVP) to optimize the MPS representation of the ground state. The reason for this is that the eigensolver they used in the DMRG calculations required the representation of not just the right eigenvector of the projected Hamiltonian, Eq.~\eqref{eq:projected Hamiltonian}, in the left-onsite-right basis, but the left eigenvector as well which has been shown to have a less compact representation \cite{Baiardi-Reiher-2020,Dobrautz-Luo-Alavi-2019}. TDVP on the other hand, does not require constructing eigenvectors of any sort, but it is much more expensive than DMRG; for the same $6 \times 6$ system and $m = 1024$ a sweep of two-site TDVP takes five times as long as a sweep of DMRG. Luckily, there has been work on developing non-Hermitian eigensolvers which only construct the right eigenvector \cite{Caricato-Trucks-2010}, and this approach was demonstrated to work for transcorrelated electronic structure DMRG calculations \cite{Liao-Zhai-Christlmaier-2023}. By using such an eigensolver we were also able to avoid using TDVP. Though not shown, we reran the pDMRG calculations from Fig.~\ref{fig:6x6-dmrg-comparison-both} with the four sweeps of DMRG replaced by four sweeps of two-site TDVP with an imaginary time step of $-0.4$, the optimal value found in Ref.~\cite{Baiardi-Reiher-2020}; neither the energy nor variance differed from the standard pDMRG calculation. We also tried augmenting pDMRG with the Krylov expansion, but again this did not produce results different from using kDMRG alone.

The final piece of our DMRG procedure is the initial state. For the systems at half-filling, for which we use the bipartite mapping, we use Eq.~\eqref{eq:bipartite k pi combo} to construct a ground state of the interaction term adapted to the $\V{K} = 0$ sector as an MPS of bond dimension 8. For the dilute systems, we use an equal superposition of free electron solutions and random product states. In the first DMRG iteration at a bond dimension of eight, we do four sweeps of pDMRG followed by four sweeps of DMRG. Then we follow the kDMRG procedure outlined above to grow the bond dimension. Though not exemplified in the system shown here, we found cases where the use of the perturbation at the smallest bond dimension produced better results.

\section{Transcorrelated bond dimension} \label{sec:Transcorrelated bond dimension}

\begin{figure}
  \includegraphics[scale=0.55]{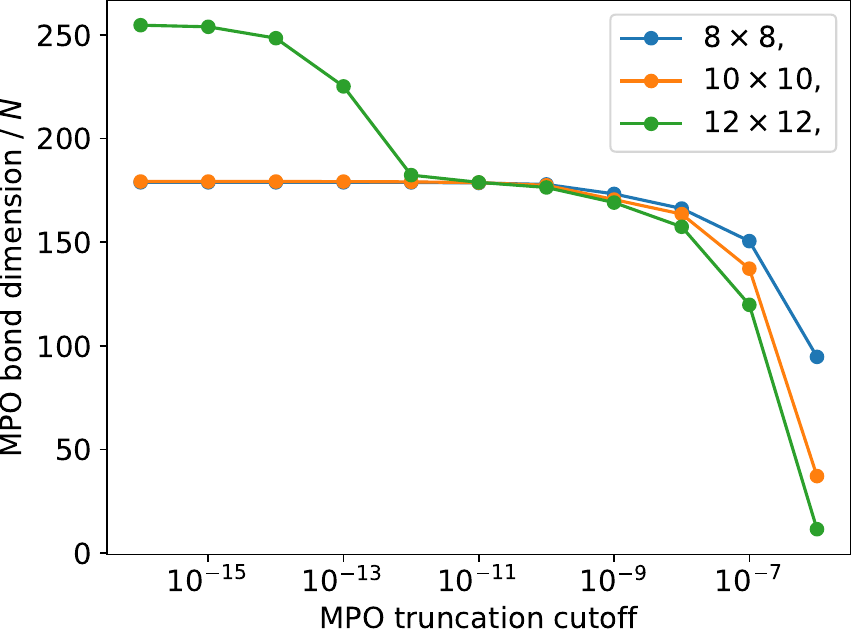}
  \caption{\label{fig:mpo-truncation} The bond dimension of the transcorrelated Hamiltonian with the bipartite mapping and $J = -0.5$ after truncation for different systems, normalized by the system size $N$.}
\end{figure}

As shown in Fig.~\ref{fig:mpo-bond-dims}, the transcorrelated Hamiltonian for the $12 \times 12$ system has an anomalously large bond dimension. We claim that this is an artifact of the numerical noise present in constructing an MPO for an operator with $3 \times 10^{10}$ terms using 143 consecutive QR decompositions. If so, redoing the construction with a larger QR tolerance should bring the bond dimension back in line with the smaller systems. The time required for this reconstruction was prohibitive, so instead we took the already constructed MPO and truncated the bond dimension using the SVD. We did this not only for the $12 \times 12$ system, but for the $8 \times 8$ and $10 \times 10$ systems as well. We swept the truncation parameter $\delta$, which corresponds to the maximum sum of the squares of the discarded singular values at each site, from $10^{-16}$ to $10^{-6}$. In Fig.~\ref{fig:mpo-truncation}, we plot the bond dimensions obtained after each truncation sweep, normalized by the system size $N$. In all cases the bond dimension decreases rapidly for $\delta > 10^{-9}$, as presumably valuable components of the Hamiltonian are being thrown away. For $\delta \leq 10^{-9}$ the bond dimensions of the $8 \times 8$ and $10 \times 10$ MPOs remain mostly unchanged, implying that they have few small singular values to discard. The bond dimension of the $12 \times 12$ MPO however, begins to decrease at $\delta = 10^{-13}$, and for $10^{-12} \leq \delta \leq 10^{-9}$ the bond dimension exhibits the same linear scaling as the smaller systems. Using the $12 \times 12$ MPO truncated with $\delta = 10^{-12}$, we verified that the transcorrelated energies remained unchanged to six figures. This strongly suggests that the extraneous bond dimension for this system is due to numerical noise.

\bibliography{references.bib}

\end{document}